\pdfoutput=1

\documentclass[11pt,twoside,a4paper,cmspaper,final,collab]{cms-tdr}
\def\svnVersion{c40dc1c-D}\def\svnDate{2019/08/29}\def\cmsCernNoTag{CERN-EP-2018-271}\def\cmsCernDate{\today}\def\cmsMessage{"Published in Physics Letters B as \href{http://dx.doi.org/10.1016/j.physletb.2019.134826}{\doi{10.1016/j.physletb.2019.134826}.}"}

\begin{document}\cmsNoteHeader{FSQ-16-012}

\hyphenation{had-ron-i-za-tion}
\hyphenation{cal-or-i-me-ter}
\hyphenation{de-vices}
\RCS$HeadURL$
\RCS$Id$
\newlength\cmsFigWidth
\ifthenelse{\boolean{cms@external}}{\setlength\cmsFigWidth{0.85\columnwidth}}{\setlength\cmsFigWidth{0.4\textwidth}}
\ifthenelse{\boolean{cms@external}}{\providecommand{\cmsLeft}{top\xspace}}{\providecommand{\cmsLeft}{left\xspace}}
\ifthenelse{\boolean{cms@external}}{\providecommand{\cmsRight}{bottom\xspace}}{\providecommand{\cmsRight}{right\xspace}}
\newlength\cmsTabSkip\setlength{\cmsTabSkip}{1ex}
\ifthenelse{\boolean{cms@external}}{\providecommand{\cmsTable}[1]{#1}}{\providecommand{\cmsTable}[1]{\resizebox{\textwidth}{!}{#1}}}
\providecommand\Fig[1]{Fig.\,\ref{fig:#1}}
\providecommand{\str}{{\sc starlight}}
\providecommand{\gaga}{\ensuremath{\gamma\gamma}}
\providecommand{\mgg}{\ensuremath{m^{\gaga}}}
\providecommand{\ee}{\ensuremath{\Pep\Pem}}
\providecommand{\Aco}{\ensuremath{\text{A}_\phi}}
\cmsNoteHeader{FSQ-16-012}
\title{Evidence for light-by-light scattering and searches for axion-like particles in ultraperipheral PbPb collisions at \texorpdfstring{$\sqrtsNN = 5.02\TeV$}{sqrt(s[NN]) = 5.02 TeV}}
\author{The CMS Collaboration}

\date{\today}

\abstract{Evidence for the light-by-light scattering process, $\gaga\to\gaga$,
in ultraperipheral PbPb collisions at a centre-of-mass energy per nucleon pair of 5.02\TeV is reported.
The analysis is conducted using a data sample corresponding to an integrated luminosity of 390\mubinv recorded
by the CMS experiment at the LHC. Light-by-light scattering processes are selected in events with
two photons exclusively produced, each with transverse energy $\et^{\gamma}>2\GeV$, pseudorapidity $\abs{\eta^{\gamma}}<2.4$,
diphoton invariant mass $\mgg> 5\GeV$, diphoton transverse momentum $\pt^{\gaga}<1\GeV$, and
diphoton acoplanarity below 0.01. After all selection criteria are applied, 14 events are observed, compared to
expectations of $9.0 \pm 0.9\thy$ events for the signal and $4.0 \pm 1.2\stat$
for the background processes. The excess observed in data relative to the background-only expectation corresponds to a
significance of 3.7 standard deviations, and has properties consistent with those expected for the light-by-light
scattering signal. The measured fiducial light-by-light scattering cross section,
$\sigma_\text{fid} (\gaga \to \gaga)=120 \pm46\stat \pm28\syst\pm12\thy\unit{nb}$,
is consistent with the standard model prediction. The $\mgg$ distribution is used to set new exclusion limits on the
production of pseudoscalar axion-like particles, via the $\gaga \to \Pa \to \gaga$ process, in the mass range
$m_\Pa = 5$--90\GeV.
}

\hypersetup{
pdfauthor={CMS Collaboration},
pdftitle={Evidence for light-by-light scattering and searches for axion-like particles in ultraperipheral PbPb collisions at sqrt(s[NN]) = 5.02 TeV},
pdfsubject={CMS},
pdfkeywords={Light-by-light, CMS, UPC, photoproduction, PbPb}}

\maketitle

\section{Introduction}

\begin{figure*}[hbtp]
\centering
 \includegraphics[width=0.99\textwidth]{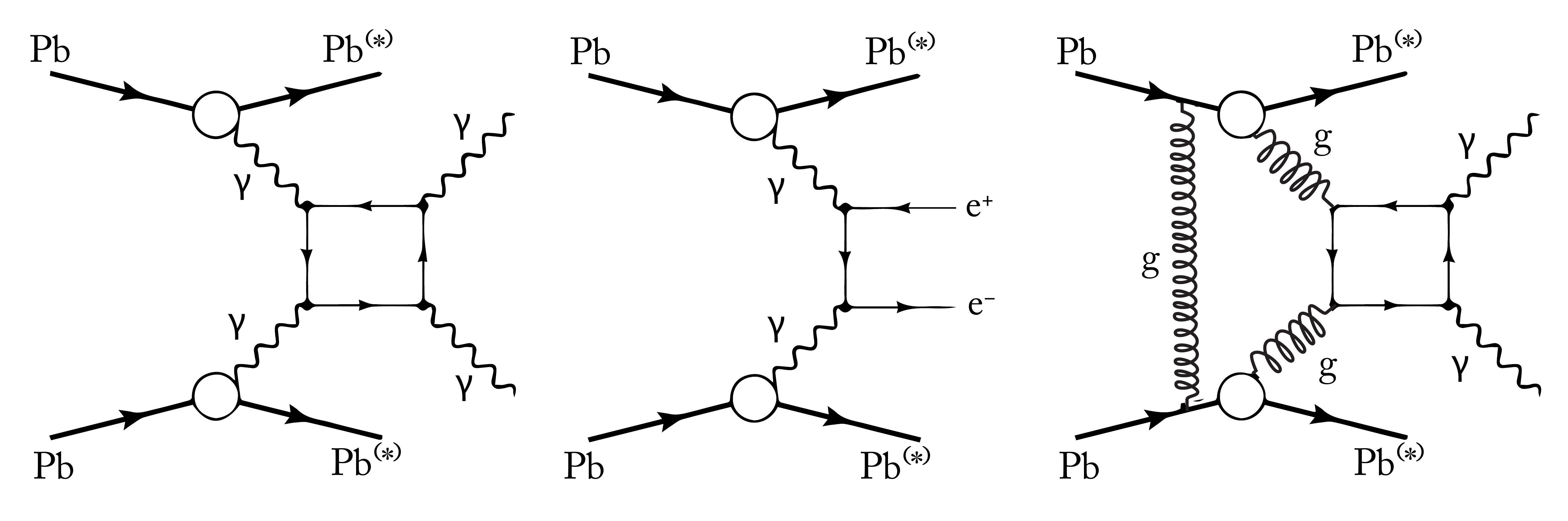}
 \caption{Schematic diagrams of light-by-light scattering ($\gaga \to \gaga$, left), QED dielectron
($\gaga \to \ee$, centre), and central exclusive diphoton ($\Pg\Pg \to \gaga$, right) production
in ultraperipheral PbPb collisions. The $\,^{(*)}$ superscript indicates a potential electromagnetic excitation
of the outgoing ions.\label{fig:feynman}}
\end{figure*}

Elastic light-by-light (LbL) scattering, $\gaga\to\gaga$, is a pure quantum mechanical process
that proceeds, at leading order in the quantum electrodynamics (QED) coupling $\alpha$, via virtual box diagrams
containing charged particles (Fig.~\ref{fig:feynman}, left). In the standard model (SM), the box diagram
involves contributions from charged fermions (leptons and quarks) and the $\PW^{\pm}$ boson.
Although LbL scattering via an electron loop has been indirectly tested through the high-precision measurements of the anomalous magnetic
moment of the electron~\cite{VanDyck:1987ay} and muon~\cite{Brown:2001mga}, its direct observation in the laboratory
remains elusive because of a very suppressed production cross section proportional to $\alpha^{4}\approx 3\times 10^{-9}$.
Out of the two closely-related processes---photon scattering in the Coulomb field of a nucleus
(Delbr\"uck scattering)~\cite{Milstein:1994zz} and photon splitting in a strong magnetic field
(``vacuum birefringence'')~\cite{Adler:1971wn,Bregant:2008yb}---only the former has been clearly observed~\cite{Jarlskog:1974tx}.
However, as demonstrated in Ref.~\cite{d'Enterria:2013yra}, the LbL process can be experimentally observed in ultraperipheral
interactions of ions, with impact parameters larger than twice the radius of the nuclei, exploiting the very large
fluxes of quasireal photons emitted by the nuclei accelerated at TeV energies~\cite{Baltz:2007kq}.
Ions accelerated at high energies generate strong electromagnetic fields, which, in the equivalent photon
approximation~\cite{vonWeizsacker:1934nji,Williams:1934ad,Fermi:1925fq}, can be considered as $\gamma$ beams of virtuality
$Q^{2} < 1/R^{2}$, where $R$ is the effective radius of the charge distribution. For lead (Pb) nuclei with radius
$R \approx 7\unit{fm}$, the quasireal photon beams have virtualities $Q^{2} < 10^{-3} \GeV^2$, but very large
longitudinal energy (up to $E_\gamma = \gamma/R \approx 80\GeV$, where $\gamma$ is the Lorentz relativistic factor), enabling
the production of massive central systems with very soft transverse momenta ($\pt\lesssim 0.1 \GeV$). Since each photon
flux scales as the square of the ion charge $Z^{2}$, $\gaga$ scattering cross sections in PbPb collisions are enhanced
by a factor of $Z^{4}\simeq 5\times 10^{7}$ compared to similar proton-proton or electron-positron interactions.

Many final states have been measured in photon-photon interactions in ultraperipheral collisions of proton
and/or lead beams at the CERN LHC, including $\gaga\to\ee$~\cite{Chatrchyan:2011ci,Chatrchyan:2012tv,
Cms:2018het,Abelev:2012ba,Abbas:2013oua,TheALICE:2014dwa,Aad:2015bwa,Aaboud:2017oiq,Aaij:2013jxj,Aaij:2014iea,Cms:2018het},
$\gaga\to \PW^+\PW^-$~\cite{Chatrchyan:2013akv,Khachatryan:2016mud,Aaboud:2016dkv}, and first evidence of
$\gaga\to\gaga$ reported by the ATLAS experiment~\cite{Aaboud:2017bwk} with a signal significance of 4.4 standard deviations (3.8 standard deviations expected).
The final-state signature of interest in this analysis is the exclusive production of two photons,
$\text{PbPb}\to\gaga\to \text{Pb}^{(*)}\gaga \text{Pb}^{(*)}$, where the diphoton final state is measured in the
otherwise empty central part of the detector, and the outgoing Pb ions (with a potential electromagnetic excitation
denoted by the $^{(*)}$ superscript) survive the interaction and escape undetected  at very low $\theta$ angles with
respect to the beam direction (Fig.~\ref{fig:feynman}, left). The dominant backgrounds are the QED production of an exclusive electron-positron pair ($\gaga\to\,\Pe^+\Pe^-$)
where the $\Pe^\pm$ are misidentified as photons (Fig.~\ref{fig:feynman}, centre), and gluon-induced
central exclusive production (CEP)~\cite{Khoze:2004ak} of a pair of photons (Fig.~\ref{fig:feynman}, right).

The $\gaga\to\gaga$ process at the LHC has been proposed as a particularly sensitive channel to study physics beyond the SM.
Modifications of the LbL scattering rates can occur if, \eg{} new heavy particles, such as magnetic monopoles~\cite{Ginzburg:1998vb},
vector-like fermions~\cite{Fichet:2014uka}, or dark sector particles~\cite{Fichet:2016clq}, contribute to the virtual corrections of
the box depicted in Fig.~\ref{fig:feynman}. Other new spin-even particles, such as axion-like particles (ALPs)~\cite{Knapen:2016moh}
or gravitons~\cite{Sun:2014qba}, can also contribute to the LbL scattering continuum or to new diphoton resonances.
In addition, light-by-light cross sections are sensitive to Born--Infeld extensions of QED~\cite{Ellis:2017edi}, and anomalous
quartic gauge couplings~\cite{Chapon:2009hh}.

We report a study of the $\gaga\to\gaga$ process, using PbPb collision data recorded by the CMS experiment
in 2015 at $\sqrtsNN = 5.02\TeV$ and corresponding to an integrated luminosity of 390\mubinv.
A comparison of exclusive diphoton and dielectron yields, with almost identical
event selection and reconstruction efficiencies, is provided as a function of key kinematic variables to check of the robustness of the analysis.
The ratio of the LbL scattering to QED $\ee$ production cross sections is reported, so as to reduce the
dependence on various experimental corrections and uncertainties. Using the measured $\mgg$ distribution, new exclusion limits are set on
ALP production, in the mass range $m_\Pa = 5$--90\GeV.

\section{The CMS detector}

The central feature of the CMS apparatus is a superconducting solenoid of 6\unit{m} internal diameter, providing a magnetic field of
3.8\unit{T}. Within the solenoid volume are a silicon pixel and strip tracker, a lead tungstate crystal
electromagnetic calorimeter (ECAL), and a brass and scintillator hadron calorimeter (HCAL), each composed of a barrel (EB and HB)
and two endcap (EE and HE) sections. Forward calorimetry (HF), based on a steel absorber and quartz fibres that run
longitudinally through the absorber and collect Cherenkov light, primarily from the electromagnetic particles,
complements the coverage provided by the barrel and endcap detectors up to pseudorapidity $\abs{\eta} = 5.2$.
Muons are measured in gas-ionisation detectors embedded in the steel
flux-return yoke outside the solenoid. The silicon tracker measures charged particles within the
$\abs{\eta}< 2.5$ range. It consists of 1440 silicon pixel and 15\,148
silicon strip detector modules. For nonisolated charged particles in the transverse momentum range $1 < \pt < 10\GeV$
and $\abs{\eta} < 1.4$, the track resolutions are typically 1.5\% in \pt
and 25--90 (45--150)\mum in the transverse (longitudinal) impact parameter~\cite{Chatrchyan:2014fea}.
The first level of the CMS trigger system~\cite{Khachatryan:2016bia}, Level-1 (L1),
composed of custom hardware processors, uses information from the calorimeters and muon detectors
to select the most interesting events in a fixed time interval of less than 4\mus. The high-level
trigger (HLT) processor farm further decreases the event rate  before data storage.
A more detailed description of the CMS detector, together with a definition
of the coordinate system used and the relevant kinematic variables,
can be found in Ref.~\cite{Chatrchyan:2008zzk}.

\section{Simulation and reconstruction}
\label{mc}

The light-by-light signal is generated with the \MADGRAPH v5~\cite{MadGraph} Monte Carlo (MC) event generator,
with the modifications discussed in Refs.~\cite{dEnterria:2009cwl,d'Enterria:2013yra} to include the nuclear photon fluxes and
the elementary LbL scattering process. The latter includes all quark and lepton loops at leading order, but omits the
{\PW} boson contributions, which are only important for diphoton masses $\mgg > 2 m_{\PW}$.
Next-to-leading order (NLO) quantum chromodynamics and QED corrections increase $\sigma_{\gaga \to \gaga}$
by just a few percent~\cite{Bern:2001dg} and are also neglected here.
Exclusive $\gaga\to\ee$ events can be misidentified as LbL scattering if neither electron track is reconstructed or if
both electrons undergo hard bremsstrahlung. This QED process is generated using the \textsc{starlight}~v2.76~\cite{Starlight:2004,Klein:2016yzr}
event generator, also based on the equivalent-photon fluxes. Since the cross section for the QED $\ee$ background is four
to five orders of magnitude larger than that for LbL scattering, and it relies on physics objects
(electrons) that are very similar to those of the signal (photons), the exclusive dielectron background is analysed in depth in order
to estimate many of the (di)photon efficiencies directly from the data, as well as to determine an LbL/(QED $\ee$) production cross sections
ratio with reduced common uncertainties. The central exclusive production process, $\Pg\Pg\to\gaga$, is simulated using
\textsc{superchic 2.0}~\cite{Harland-Lang:2015cta}, where the computed proton-proton cross section~\cite{Khoze:2004ak} is conservatively scaled to the
PbPb case by multiplying it by $A^{2}R_{\Pg}^{4}$, where $A = 208$ is the mass number of lead and $R_{\Pg} \approx 0.7$ is a gluon shadowing correction
in the relevant kinematic range~\cite{Eskola:2016oht}, and where the rapidity gap survival probability, encoding the probability to produce
the diphoton system exclusively without any other hadronic activity, is assumed to be 100\%. Given the large theoretical uncertainty of the
CEP process for PbPb collisions, the absolute normalisation of this MC contribution is directly determined from a control region in
the data, as explained later.
All generated events are passed through the \GEANTfour~\cite{Agostinelli:2002hh} detector simulation, and the events are
reconstructed with the same software as for collision data.
The simulation describes the tracker material budget with an accuracy better than 10\%, as established by measuring the
distribution of reconstructed nuclear interactions and photon conversions in the tracker~\cite{Chatrchyan:2014fea, Sirunyan:2018icq}.

Photons and electrons are reconstructed using an algorithm based on the particle flow global event description (GED)~\cite{Sirunyan:2017ulk}.
The GED algorithm uses information from each subdetector system to provide charged-particle tracks, calorimeter clusters, and muon tracks.
Electromagnetic showers from photons and electrons deposit 97$\%$ of their incident energy into an array of $5{\times}5$ ECAL crystals.
The tracker material can induce photon conversion and electron bremsstrahlung and, because of the presence of the strong CMS solenoidal magnetic
field, the energy reaching the calorimeter is thereby spread in $\phi$. The spread energy is captured by building
a cluster of clusters, or ``supercluster''~\cite{Khachatryan:2015iwa}. The GED algorithm allows for an almost complete recovery of the energy
of the photons and electrons, even if they initiate an electromagnetic shower in the material
in front of the ECAL. Nonetheless, in the case of photons, in order to keep to a minimum the $\Pe^{\pm}$ contamination,
we require them to be unconverted in the tracker. The reconstructed energy of this supercluster is used to define the energy of the photon.
Since the default CMS photon reconstruction algorithm is optimised for $\gamma$ and $\Pe^\pm$ with transverse energies $\et = E \sin \theta > 10\GeV$,
whereas the cross section for photons and electrons from exclusive production peaks in the lower $\et \approx 2\textrm{--}10\GeV$ range,
a version of the GED algorithm optimised for this transverse energy range is employed.
The threshold for the energy of photons, electrons, and superclusters, is lowered to
$1\GeV$, instead of the 10--15\GeV threshold used in the standard CMS analyses~\cite{Khachatryan:2015iwa}.
The full analysis is independently repeated using a different ``hybrid'' photon/electron reconstruction algorithm~\cite{Khachatryan:2015iwa},
obtaining reconstruction efficiencies and final results fully consistent with those derived with the default GED approach.
Additional particle identification (ID) criteria are applied, in order
to remove photons (mostly) from high-\pt neutral pion decays, based on a shower shape analysis that requires the width of
the electromagnetic shower along the $\eta$ direction to be below 0.02 (0.06) units in the ECAL barrel (endcap).

Electron candidates are identified by the association of a charged-particle track from the collision vertex
with superclusters in the ECAL. The association takes into account
energy deposits both from the electron and from bremsstrahlung photons produced during its
passage through the inner detector. Additional electron identification criteria (isolation, number of tracker hits,
HCAL/ECAL energy deposit) are applied, as discussed in Ref.~\cite{Chatrchyan:2012tv}.
The electron energy scale is verified using a sample of $\gaga\to\ee$ events, comparing the energy of
the supercluster $E$ to the momentum of the track $p$. The electron $E/p$ ratio is within 5\% of unity in the barrel and 15\%
in the endcaps. A good agreement is found between data and simulation, both in the energy scale and resolution.
In addition, the LbL simulation is also used to validate the photon energy scale.
The reconstructed supercluster energy and generated photon energy agree within a few percent,
confirming that the reconstructed supercluster energy is well calibrated.

\section{Event selection and background estimation}

The exclusive diphoton candidates are selected at the trigger level with a dedicated L1 algorithm
that requires at least two electromagnetic clusters (L1 EG) with \et above 2\GeV and at least one of
the HF detectors with total energy below the noise threshold. No additional selection is applied in the HLT.
Data are also recorded with single-photon triggers with \et thresholds above 5 and 10\GeV,
and used in this analysis to estimate the efficiency of the first trigger via a tag-and-probe
procedure~\cite{Khachatryan:2010xn}, as described below.
In the offline analysis, events are selected with exactly two photons,
each with  $\et > 2\GeV$ and $\abs{\eta} <2.4$, that satisfy further selection requirements described below.
Neutral and charged exclusivity selection criteria are applied to reject events having any additional
activity over the range $\abs{\eta} < 5.2$.
First, all events with reconstructed charged-particle tracks with $\pt > 0.1\GeV$
are removed from further analysis. Second, events are required to have no activity in the calorimeters,
above energy noise thresholds (ranging between about 0.6\GeV in the barrel, to 4.9\GeV in the HF),
outside a region $\Delta \eta <  0.15$ and $\Delta \phi < 0.7$
in the barrel ($\Delta \eta <  0.15$ and $\Delta \phi < 0.4$ in the endcap) around the two photons.
The noise thresholds are determined from no- or single-bunch crossing events.
To eliminate nonexclusive backgrounds, characterised by a final state with larger
\pt and larger diphoton acoplanarities, $\text{A}_{\phi} = (1-\Delta \phi^{\gaga}/\pi)$,
than the back-to-back exclusive $\gaga$ events, the transverse momentum of the diphoton system
is required to be $\pt^{\gaga}< 1\GeV$, and the acoplanarity of the pair to be
$\text{A}_{\phi} < 0.01$. The chosen values of the pair $\pt$ and acoplanarity selections,
similar to those originally suggested in Ref.~\cite{d'Enterria:2013yra}, are motivated by previous
CMS studies of exclusive dilepton production~\cite{Chatrchyan:2011ci,Chatrchyan:2012tv,Cms:2018het}.
The two dominant exclusive background sources potentially remaining in the LbL scattering signal region,
$\gaga \to \ee$ and CEP $\Pg\Pg \to \gaga$, are studied in detail next.

\subsection{QED \texorpdfstring{$\ee$}{ee} background}

\begin{figure}[htbp]
 \centering
  \includegraphics[width=0.45\textwidth]{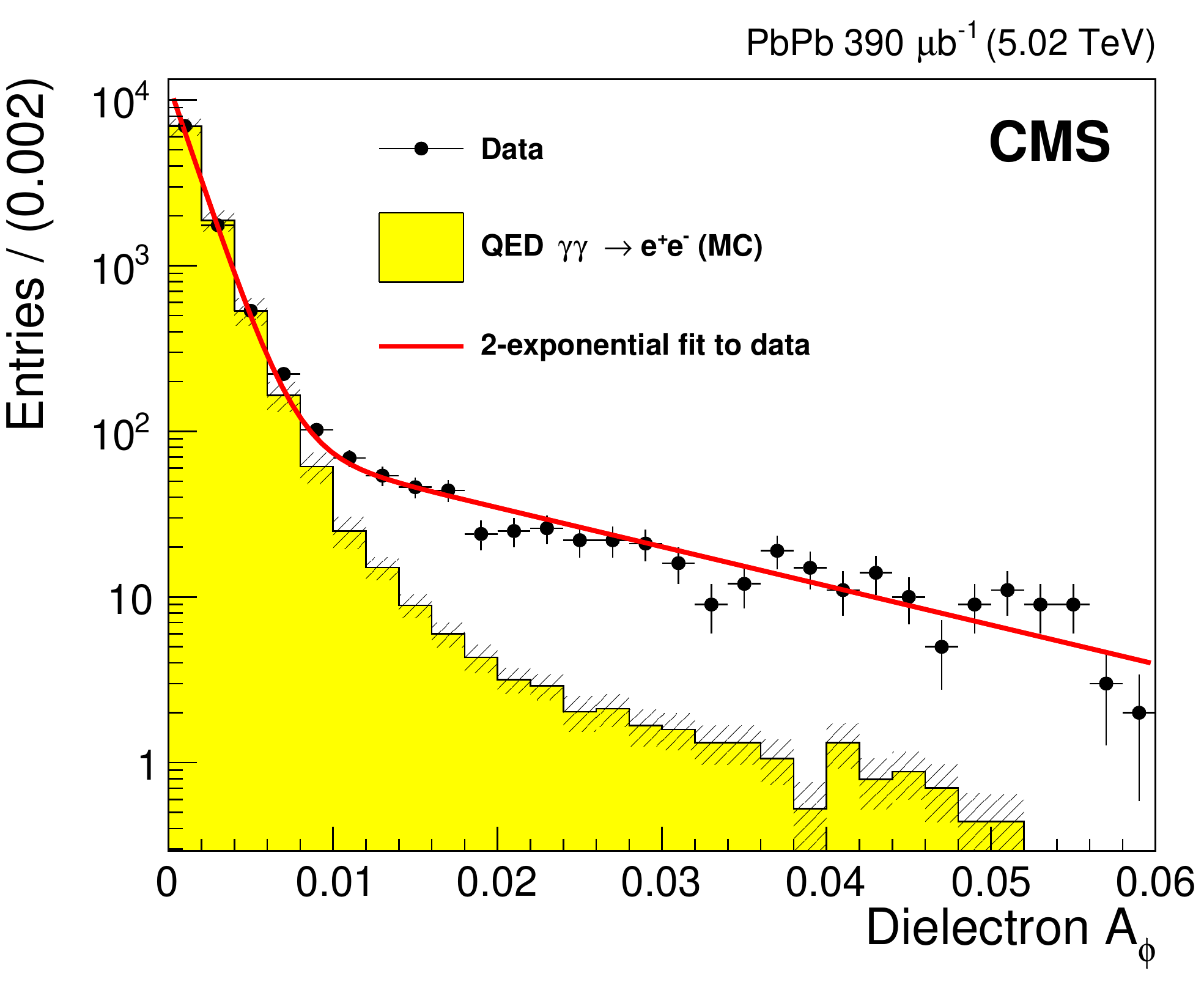}
\caption{\label{fig:qed_ee_acop} Acoplanarity distribution of exclusive $\ee$ events measured in data (circles),
compared to the expected QED $\ee$ spectrum in the \str\ MC simulation (histogram), scaled as described in the text.
The curve shows a $\chi^2$ fit to the sum of two exponential distributions corresponding to exclusive $\ee$ plus
any residual (nonacoplanar) background pairs.
Error bars around the data points indicate statistical uncertainties, and hashed bands around the histogram
include systematic and MC statistical uncertainties added in quadrature. The horizontal bars around the data symbols indicate the bin size.
}
\end{figure}

In order to have a full control of the QED $\ee$ background in the LbL scattering signal region,
the same analysis carried out for the LbL events is done on exclusive dielectron candidates,
applying the same criteria as described above for diphoton events, with the exception that exactly
two opposite-sign electrons are reconstructed, instead of exactly two photons, and no additional
track with $\pt>0.1\GeV$ should be present in addition to the two tracks corresponding to the electrons.
Figure~\ref{fig:qed_ee_acop} shows the acoplanarity distribution measured in QED $\ee$ events passing
all selection criteria compared to the MC expectation.

\begin{figure*}[!t]
 \centering
  \includegraphics[width=0.4\textwidth]{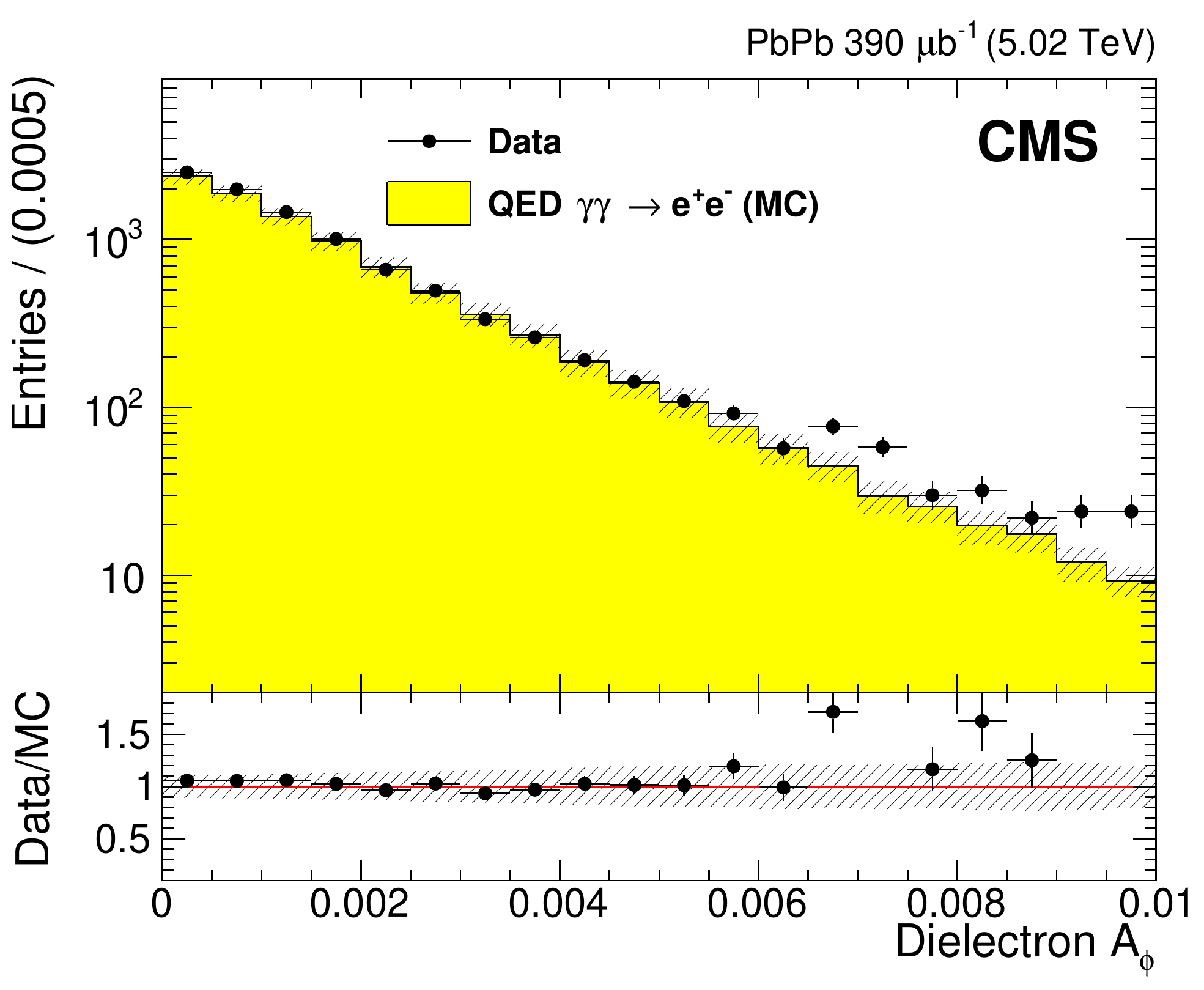}
  \includegraphics[width=0.4\textwidth]{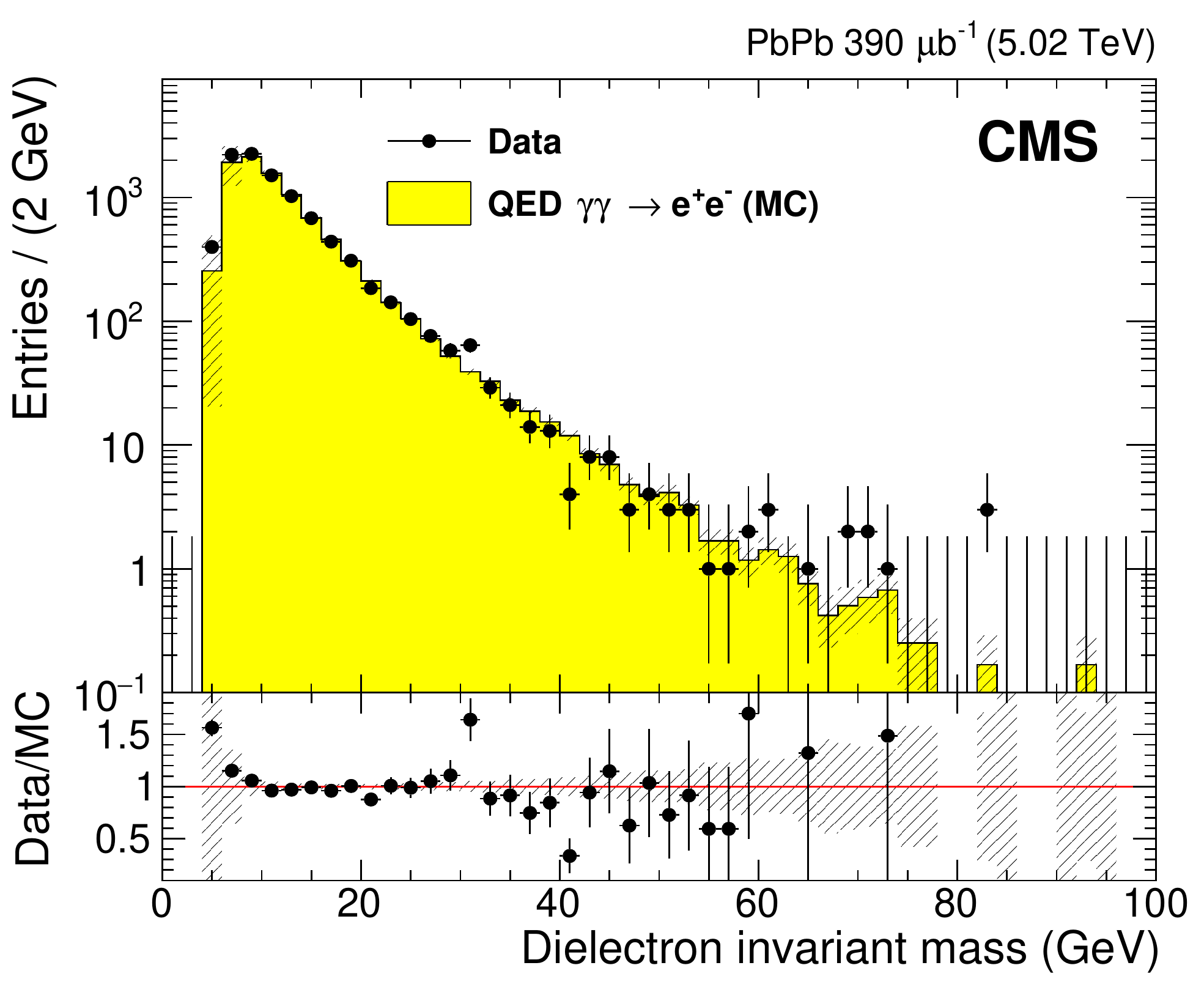}
  \includegraphics[width=0.4\textwidth]{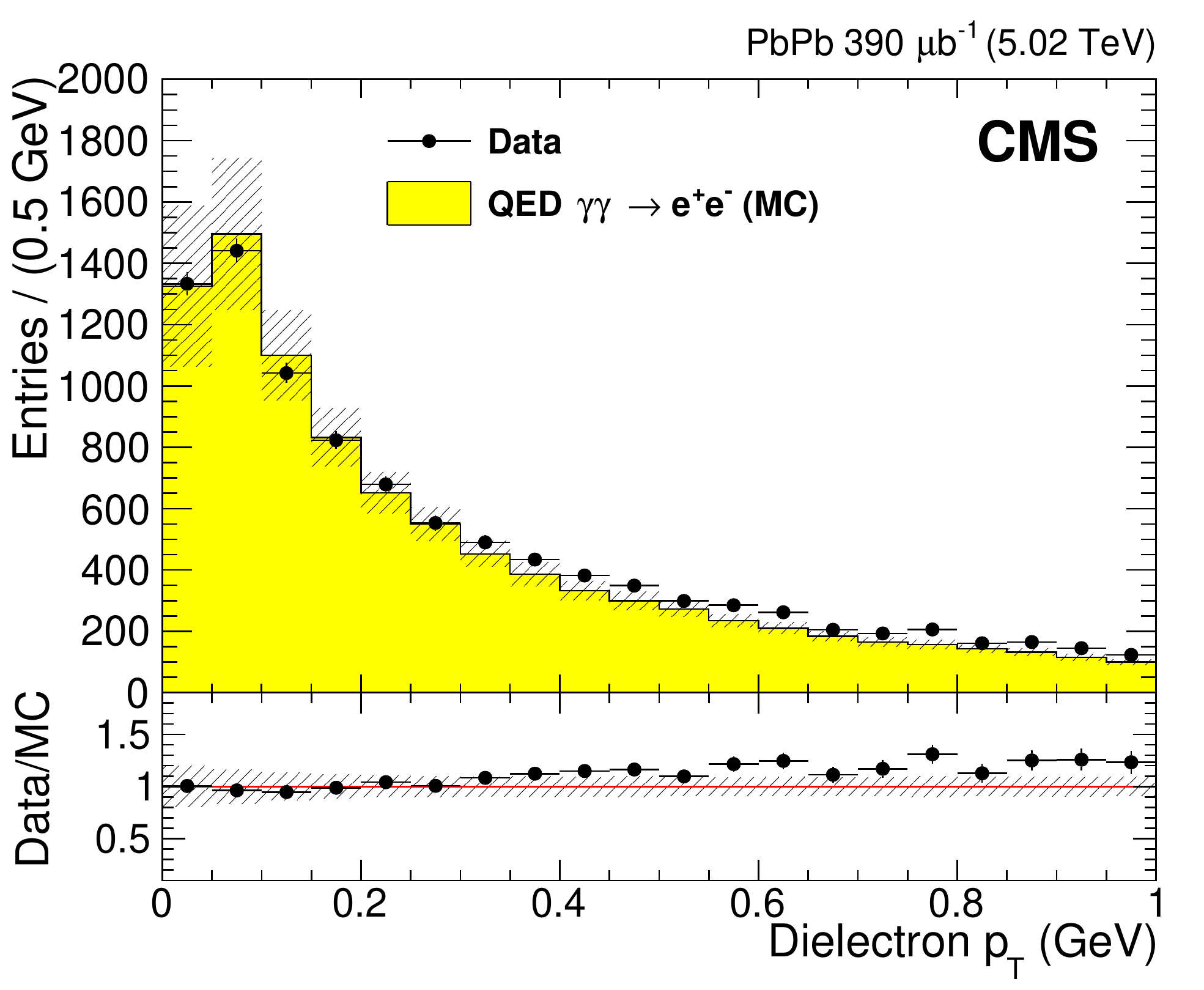}
  \includegraphics[width=0.4\textwidth]{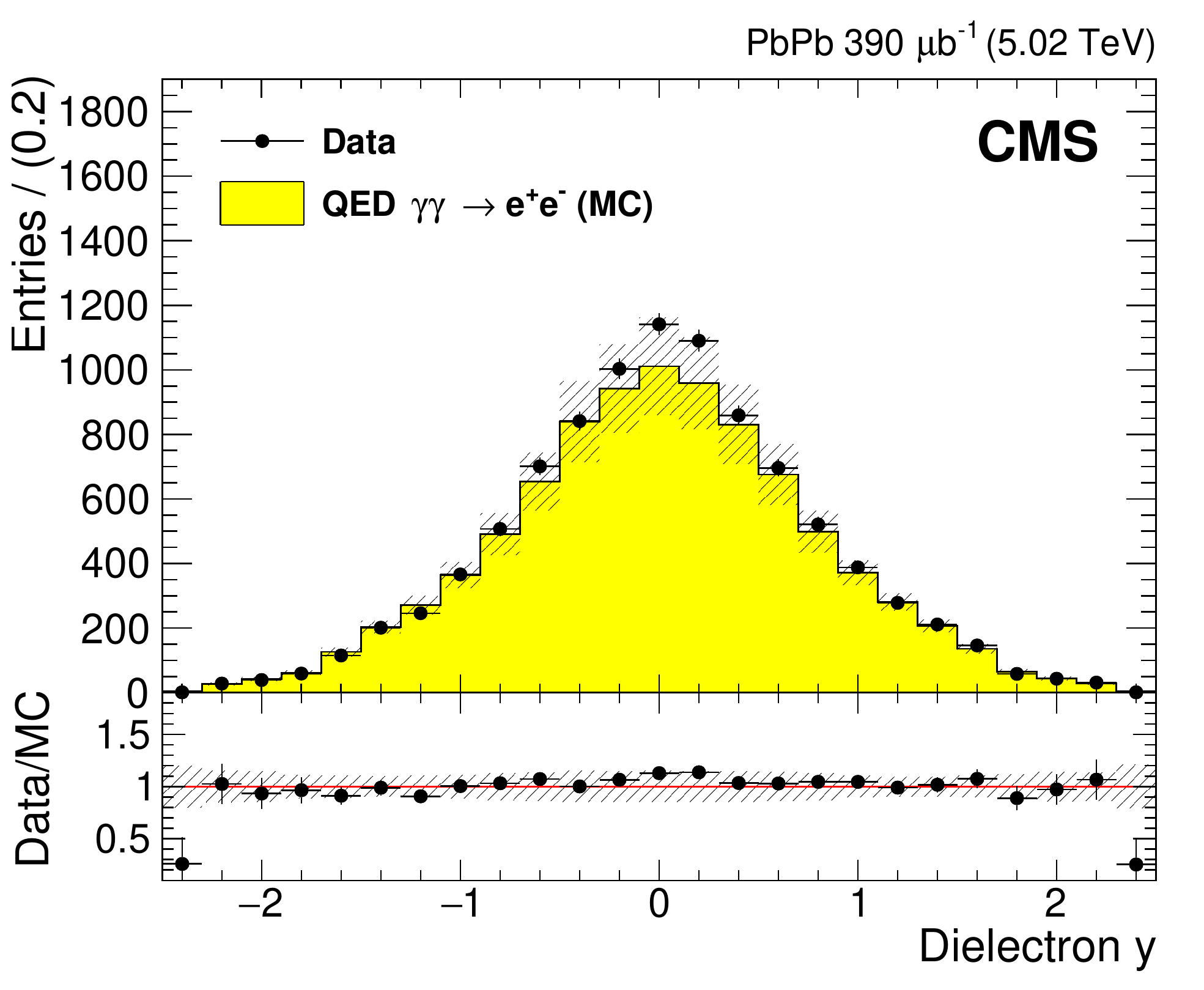}
\caption{\label{fig:qed_ee_kine} Comparison of data (circles) and \str\ MC expectation (histogram, scaled as described in the text)
for the exclusive $\ee$ events passing all selection criteria, as a function of dielectron acoplanarity (top left),  mass (top right),
\pt (bottom left), and rapidity $y$ (bottom right). Error bars around the data points indicate
statistical uncertainties, and hashed bands around the histograms include systematic and MC
statistical uncertainties added in quadrature. The horizontal bars around the data symbols
indicate the bin size. The ratio of the data to the MC expectation is shown in the bottom panels.}
\end{figure*}

The curve is a binned $\chi^2$ fit of the data to the sum of two exponential functions representing
the exclusive QED $\ee$ production plus any residual background in the high-acoplanarity tail. In the region of acoplanarity below 0.01,
9570 dielectron events are reconstructed with a purity of $\mathcal{P}=0.960 \pm 0.002\stat$, obtained
from the ratio of amplitudes of the two exponential functions fitted to the data.
The yellow histogram shows the same distribution obtained directly from the \str\ MC simulation, scaled
to the total number of events in data, multiplied by the purity.
The corresponding kinematic distributions of the selected $\gaga\to \ee$ events in the $\text{A}_{\phi} < 0.01$ region
are shown in \Fig{qed_ee_kine}, together with the corresponding MC predictions normalised in the same manner.
The hashed band around the MC histograms include the systematic uncertainties (trigger, electron reconstruction and identification, and MC
statistical uncertainties added in quadrature) discussed in Section~\ref{sec:syst}, estimated as a function of
electron \et and $\eta$. A good data-to-simulation agreement is found, thereby confirming the quality of the electromagnetic
particle reconstruction, and of the exclusive event selection criteria, as well as of the MC
predictions~\cite{d'Enterria:2013yra,Klein:2016yzr} for exclusive particle production in ultraperipheral PbPb collisions at the LHC.
Small systematic differences between the central values of the exclusive dielectron data and the MC prediction are seen
in tails of some of the distributions (at increasing acoplanarity and \pt) due to the presence of slightly acoplanar events in data,
likely from $\gaga\to\ee$ events where one (or both) electrons radiate an extra soft photon, that are not explicitly
simulated by the MC event generator. These small discrepancies have no impact on the final extracted cross sections
integrated over the whole range of distribution(s).

The QED dielectron background is then directly estimated from the \str\ MC simulation by counting the number
of such $\ee$ events that pass all LbL scattering selection criteria. The charged exclusivity condition, requiring no track
in the event above the $\pt = 0.1$\GeV threshold, is successful in removing this background almost entirely.
This tracking efficiency is controlled in events with exactly two reconstructed photons and exactly one track, finding good data-MC agreement.
The QED background in the signal region is estimated to be $N^{\Pe\Pe,\text{data}} = 1.0 \pm 0.3\stat$ events, where the assigned
uncertainty corresponds to the event count in the simulated samples.

\subsection{Central exclusive diphoton background}
\begin{figure}[hbtp]
\centering
 \includegraphics[width=0.45\textwidth]{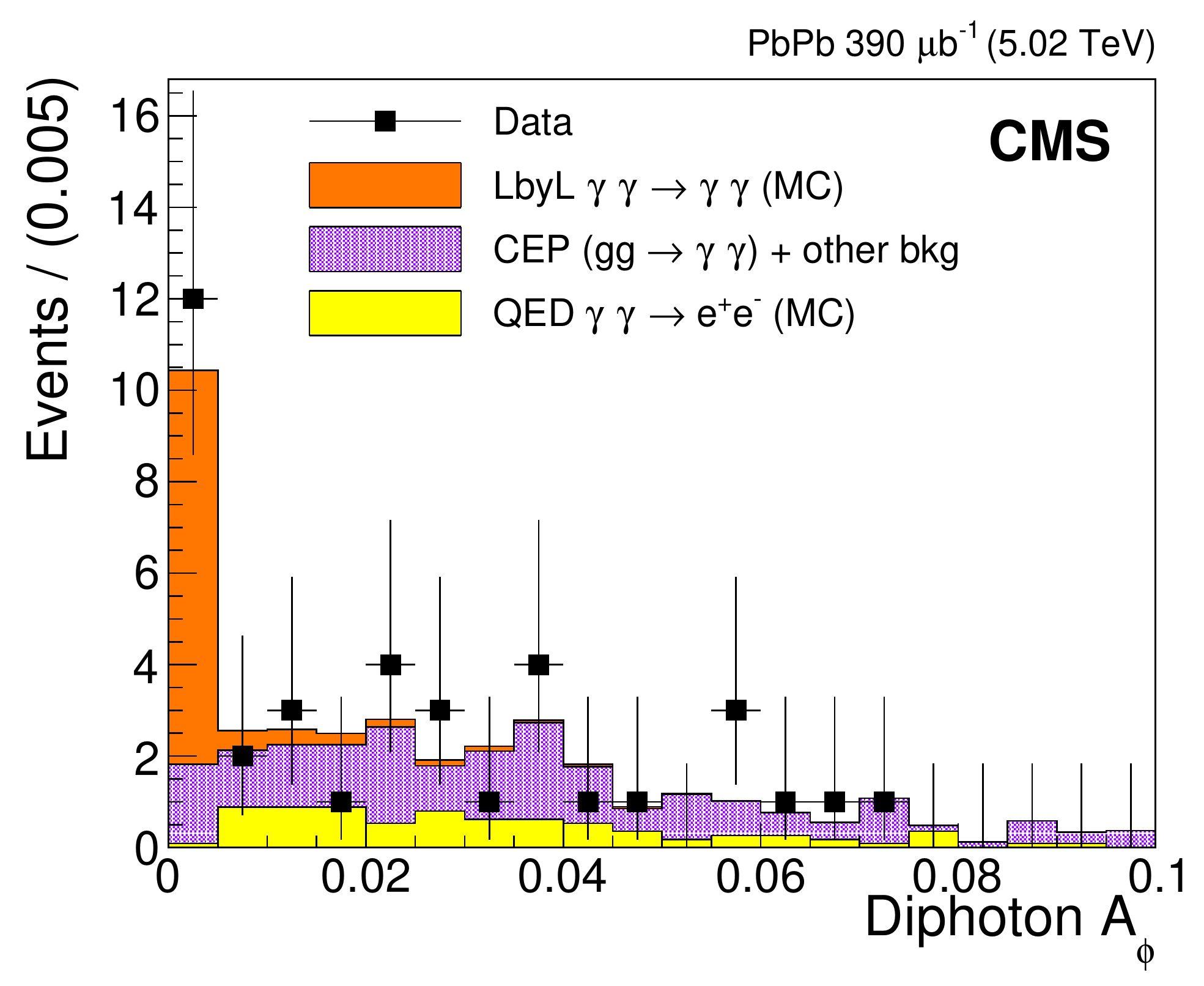}
  \caption{Diphoton acoplanarity distribution for exclusive events measured in the data after
selection criteria (squares), compared to the expected LbL scattering signal (orange histogram), QED $\ee$ (yellow histogram),
and the CEP+other (light blue histogram, scaled to match the data in the $\text{A}_{\phi} > 0.02$ region as described in the text) backgrounds.
Signal and QED $\ee$ MC samples are scaled according to their theoretical cross sections and integrated luminosity.
The error bars around the data points indicate statistical uncertainties. The horizontal bars around the data symbols indicate the bin size. }
\label{fig:aco_plot}
\end{figure}

Although the LbL and CEP processes share an identical final state, their kinematic distributions are different.
Diphotons from quasireal $\gaga$ fusion processes are produced almost at rest in the transverse plane and, thus,
the final-state photons are emitted back-to-back with balanced pair transverse momentum $\pt^{\gaga}\approx 0$. On the other hand,
typical CEP photon pairs are produced in diffractive-like gluon-mediated processes~\cite{Khoze:2004ak,Harland-Lang:2015cta} with larger momentum exchanges
leading to a diphoton transverse momentum distribution peaking at $\pt\approx 0.5\GeV$, after selection criteria, and moderately
large tails in the azimuthal acoplanarity distribution. Thus, the requirement on diphoton acoplanarity ($\text{A}_{\phi} < 0.01$)
also significantly reduces the $\Pg\Pg \to \gaga$ background. Since the MC prediction for CEP $\Pg\Pg \to \gaga$
has large theoretical uncertainties, and in order to account for any other remaining backgrounds
resulting in photons that are not back-to-back (such as $\gaga \to \ee \gamma(\gamma)$ events passing the analysis selection criteria),
for which no event generator is currently available, the CEP background is normalised to match the data in the region $\text{A}_{\phi} > 0.02$,
where the contribution from $\gaga\to\gaga$ is negligible (\Fig{aco_plot}).
The background normalisation factor is then obtained from
\begin{linenomath*}
\ifthenelse{\boolean{cms@external}}{
\begin{multline}
 f_\text{nonacoplanar}^\text{norm} = \\
 \Bigl[N_\text{data}(\text{A}_{\phi} > 0.02) - N_\text{LbL}^\text{MC}(\text{A}_{\phi} > 0.02) \\
 - N_\text{QED}^\text{MC}(\text{A}_{\phi} > 0.02)\Bigr]
 \Bigg/\Bigl[N_\text{CEP}^\text{MC}(\text{A}_{\phi} > 0.02)\Bigr],
  \label{eq:photon_eff}
\end{multline}
}{
\begin{equation}
 f_\text{nonacoplanar}^\text{norm} =
 \frac{N_\text{data}(\text{A}_{\phi} > 0.02) - N_\text{LbL}^\text{MC}(\text{A}_{\phi} > 0.02) - N_\text{QED}^\text{MC}(\text{A}_{\phi} > 0.02)}{N_\text{CEP}^\text{MC}(\text{A}_{\phi} > 0.02)},
  \label{eq:photon_eff}
\end{equation}
}
\end{linenomath*}
and found to be $f_\text{nonacoplanar}^\text{norm} = 1.06 \pm 0.35\stat$. The number of events
due to CEP plus any residual backgrounds is thus estimated to be $3.0 \pm 1.1\stat$. The
statistical uncertainties quoted in both values are driven by the size of the data sample
remaining at high acoplanarities, after all selection criteria have been applied.

\subsection{Light-by-light signal distributions}

The exclusive diphoton signal is extracted after applying all selection criteria described above
and estimating the amount of residual QED $\ee$ and CEP+other backgrounds.
Table~\ref{tab:event_sel_ged} shows the number of events remaining after each selection criterion.
The main selection requirement corresponds to two photons each with $\et > 2\GeV$,
$\abs{\eta} < 2.4$ (excluding photons falling in the $\Delta\eta\approx 0.1$ gap region
between the EB and EE, $1.444<\abs{\eta}<1.566$), and diphoton invariant mass greater than 5\GeV.
The numbers of events measured in data and expected from the sum of all MC contributions in the first
two rows do not match because these selection requirements accept a fraction of nonexclusive backgrounds that are
not included in the simulation. Once the full exclusivity selection criteria are applied,
the data-to-simulation agreement is very good. We observe 14 LbL scattering candidates,
to be compared with $9.0 \pm 0.9\thy$ expected from the LbL scattering signal,
$3.0\pm 1.1\stat$ from central exclusive plus any residual diphoton backgrounds, and $1.0\pm 0.3\stat$ from misidentified QED $\ee$ events.

An extra selection criterion has been also studied by further requiring that the candidate LbL scattering events have
no signal above the noise threshold in the pixel tracker layers. This more stringent selection is sensitive to
charged particles down to ${\sim}40\MeV$, and results in a number of reconstructed LbL scattering signal
counts (and even more reduced QED backgrounds) consistent with the MC predictions. However, since the efficiency of such a
tight selection is difficult to assess from a control region in data, the default analysis is kept with
the charged-particle track $\pt > 0.1\GeV$ exclusivity requirement.

\begin{table*}[hbtp]
\centering
\topcaption{Number of diphoton candidates measured in data and expected from MC simulation for LbL scattering,
QED $\ee$ production, and from the CEP+other contributions, after each event selection step (cumulative)
described in the text. The yields of the simulated processes are scaled according to their theoretical
cross sections and the integrated luminosity of the analysed data set. The CEP+other values are normalised from the
high-acoplanarity tail with a scale factor estimated from the data as described in the text. The LbL scattering simulation uncertainty quoted
is that of the theoretical uncertainty of the prediction, whereas the uncertainties in the QED $\ee$ and CEP+others
yields are statistical.}
    \label{tab:event_sel_ged}
  \cmsTable{
    \begin{tabular}{l r c c c}
      \hline
      Selection criteria                             & Data    & LbL MC  & QED $\ee$ MC  & CEP MC + other \\
                                                     &         &         &      & (normalised to data) \\\hline
      Charged exclusivity                            &  648    &  $11.1\pm 1.2\thy$   & $10.3\pm 1.0\stat$  &  $24.3\pm 8.1\stat$ \\
      Neutral exclusivity                            &  108    &  $10.8\pm 1.1\thy$   & $10.1\pm 1.0\stat$  &  $23.6\pm 7.9\stat$ \\
      Diphoton \pt$ < 1$ \GeV                        &  39     &  $10.2\pm 1.1\thy$   & $7.7\pm 1.0\stat$   &  $19.5\pm 6.5\stat$  \\
      Diphoton acoplanarity $< 0.01$                 &  14     &  $9.0\pm 0.9\thy$    & $1.0\pm 0.3\stat$   &  $3.0\pm 1.1\stat$ \\\hline
      \end{tabular}
}
\end{table*}

\begin{figure*}[!t]
\centering
 \includegraphics[width=0.45\textwidth]{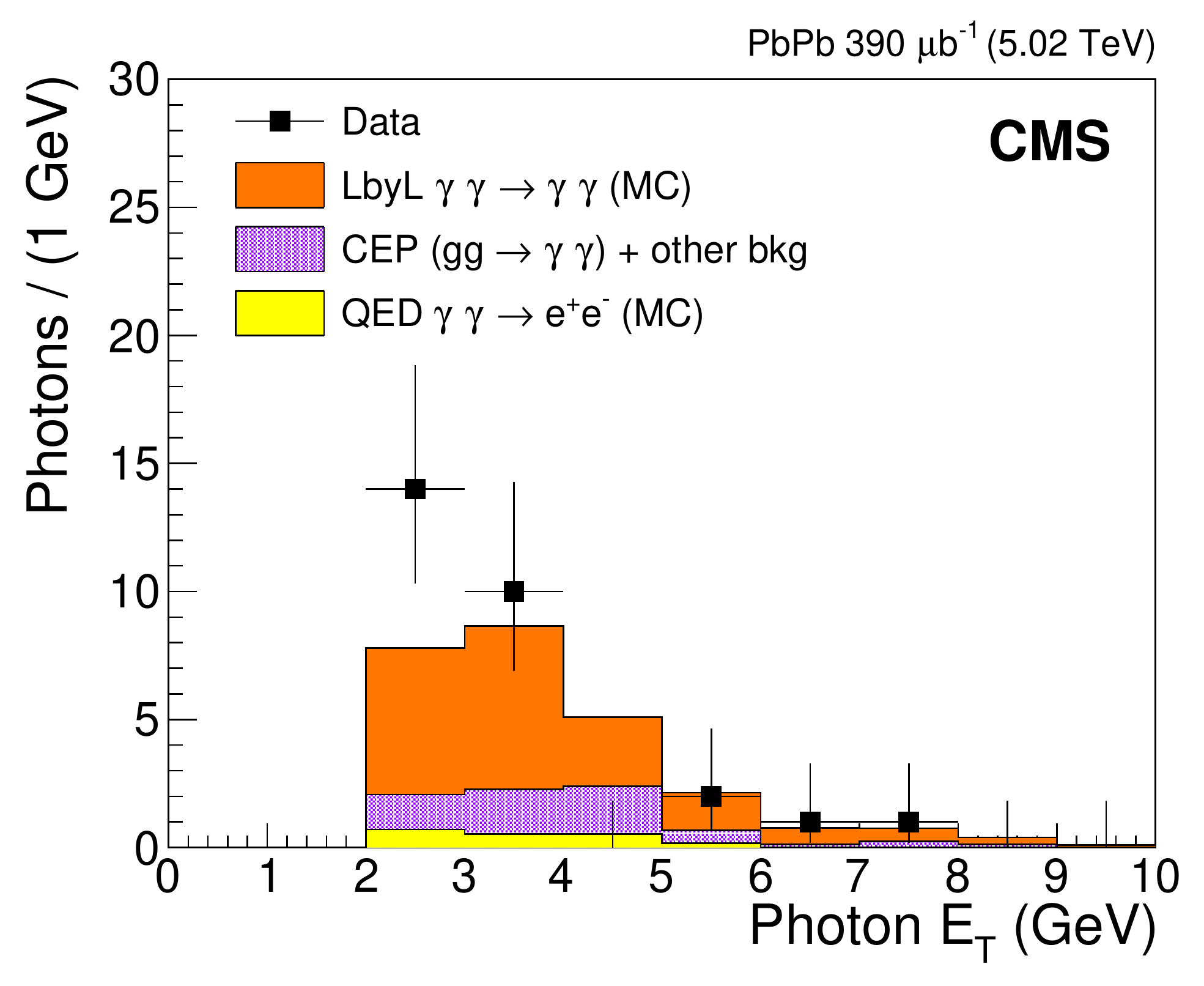}
 \includegraphics[width=0.45\textwidth]{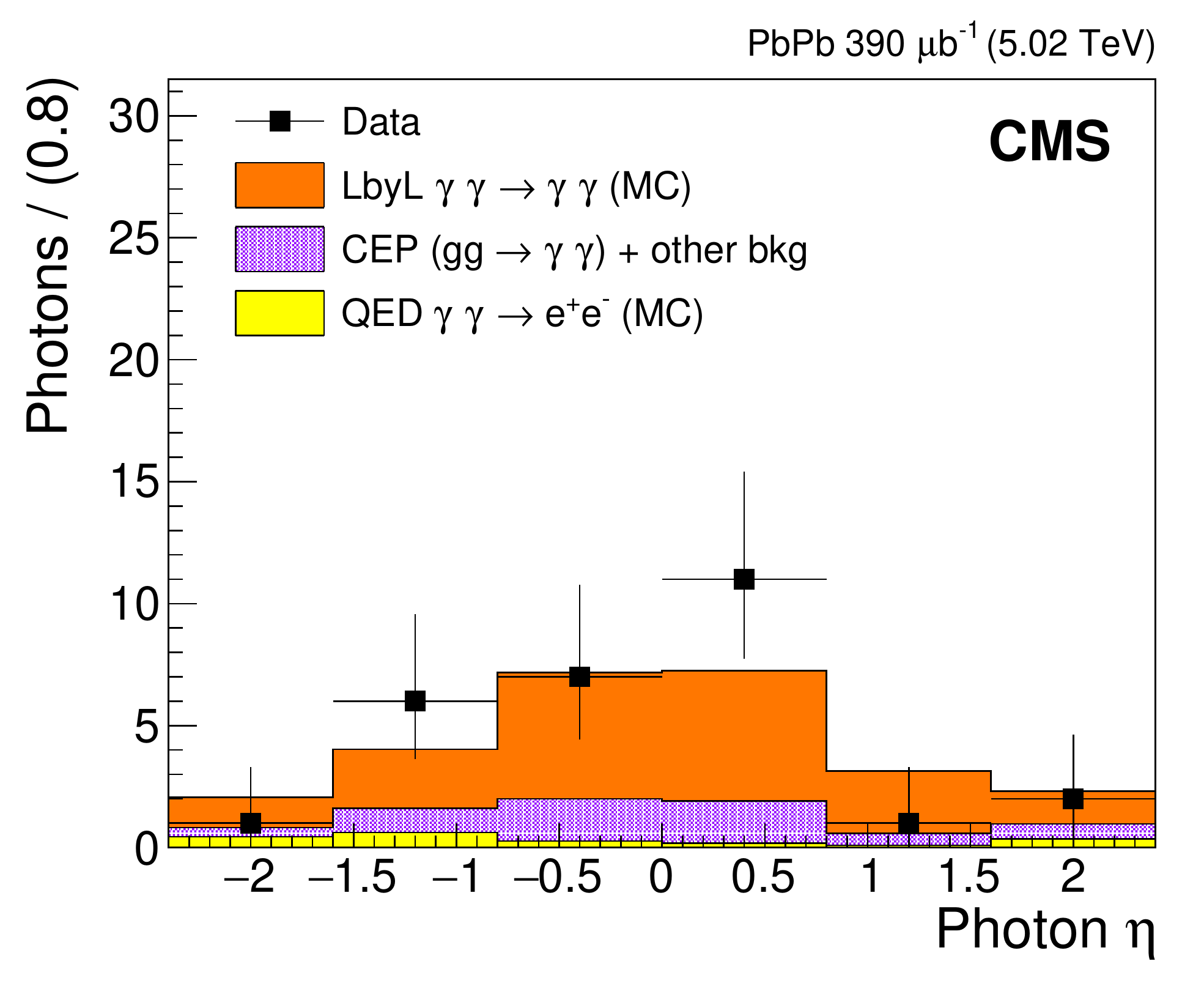}
 \includegraphics[width=0.45\textwidth]{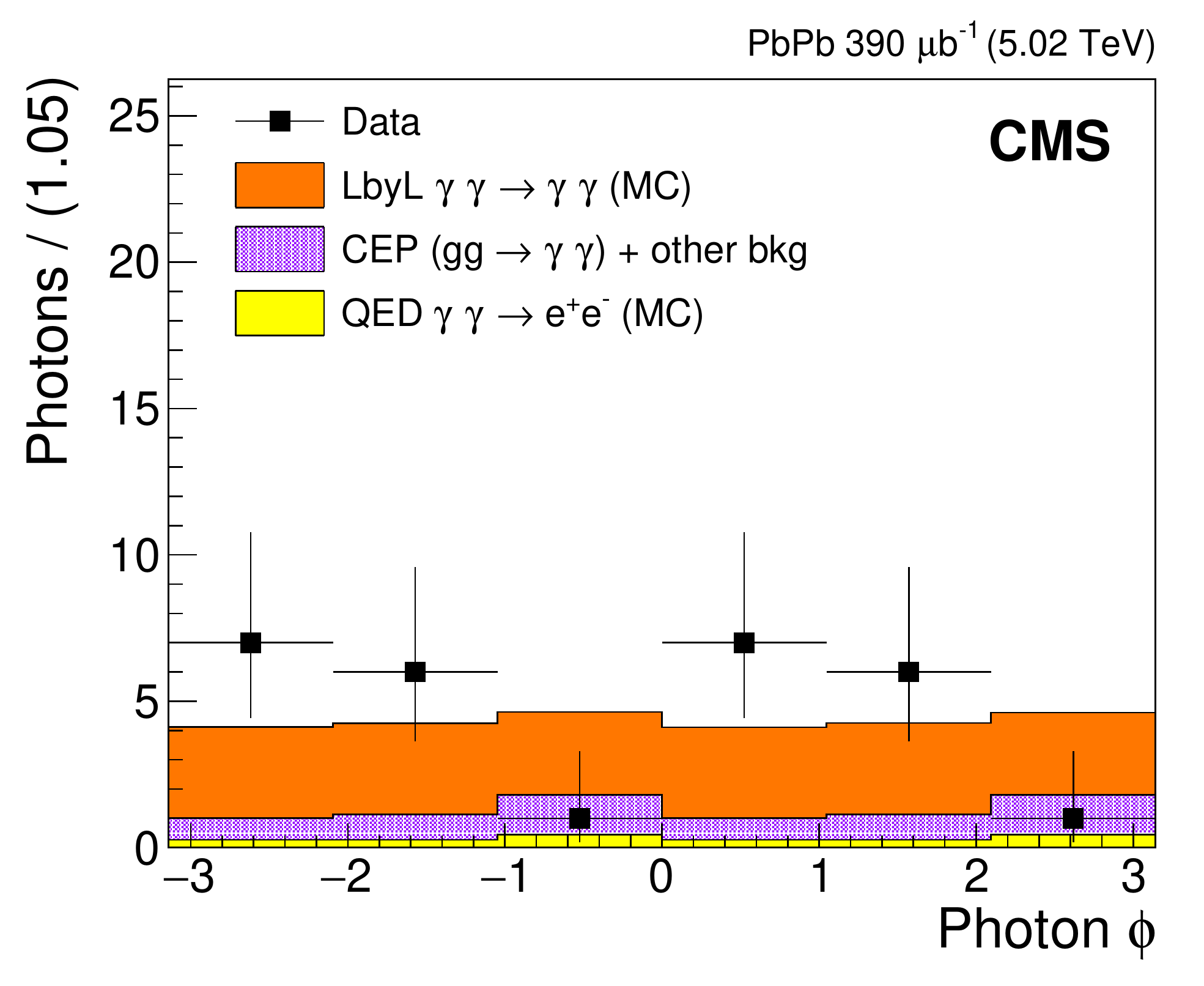}
 \includegraphics[width=0.45\textwidth]{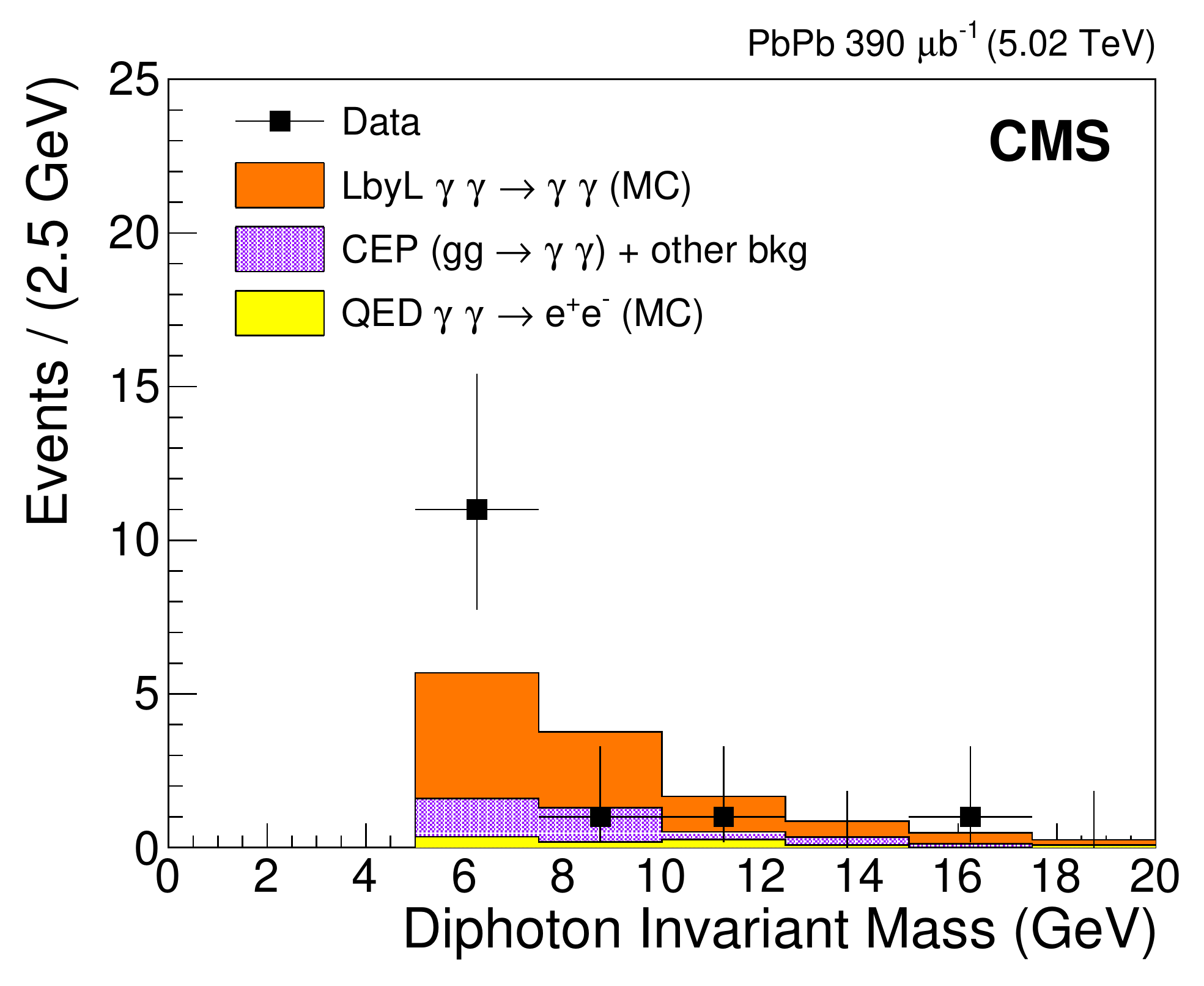}
 \includegraphics[width=0.45\textwidth]{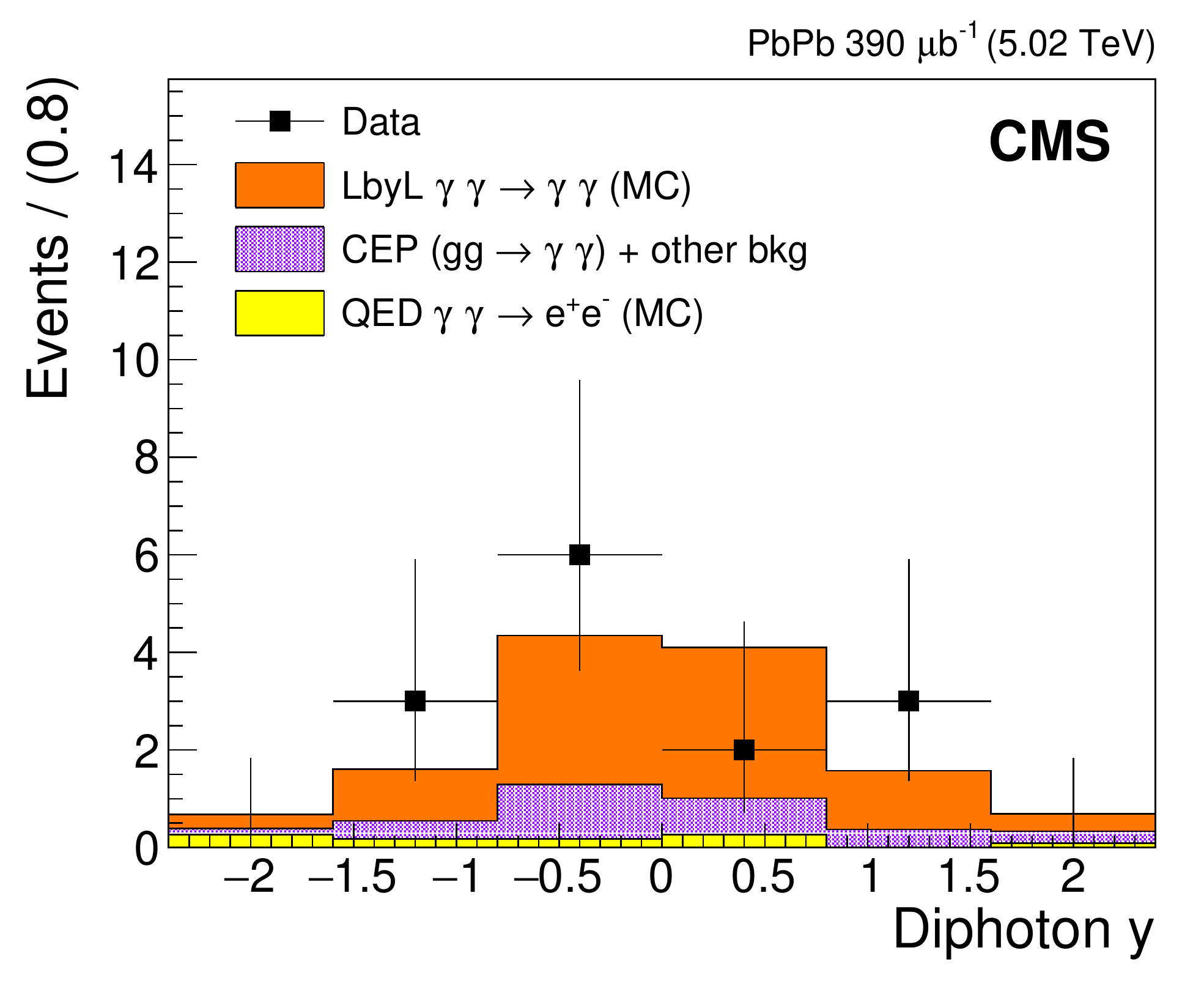}
 \includegraphics[width=0.45\textwidth]{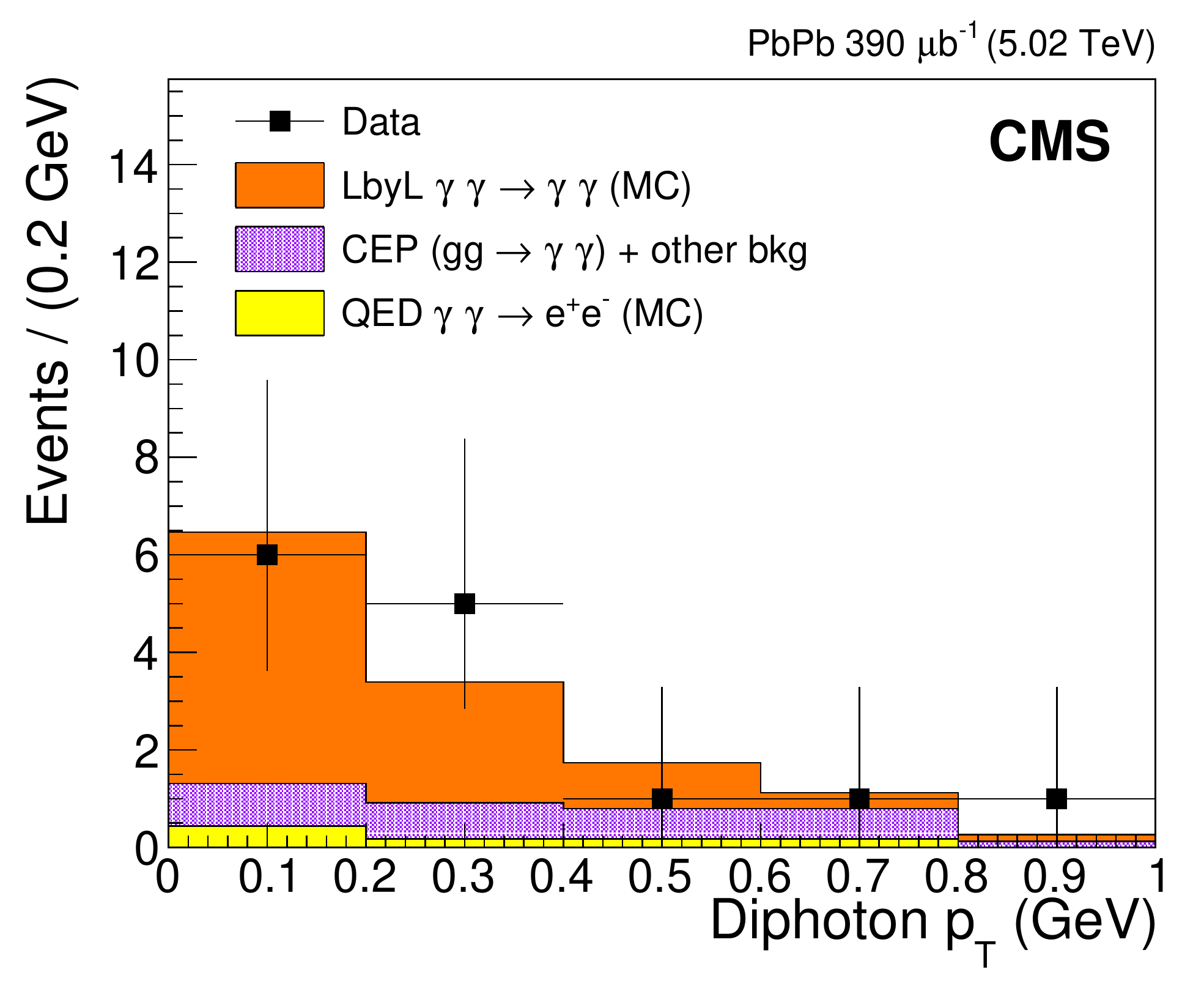}
 \caption{Distributions of the single photon \et, $\eta$, and $\phi$, as well as diphoton \pt, rapidity, and invariant mass
measured for the fourteen exclusive events passing all selection criteria (squares), compared to the expectations of LbL scattering signal (orange histogram),
QED $\ee$ MC predictions (yellow histogram), and the CEP plus other backgrounds (light blue histogram, scaled to match the data in
the $\text{A}_{\phi} > 0.02$ region). Signal and QED $\ee$ MC samples are scaled according to their theoretical cross sections and integrated
luminosity. The error bars around the data points indicate statistical uncertainties. The horizontal bars around the data symbols indicate the bin size. }
\label{fig:data_mc_ged_af_scale}
\end{figure*}

Figure~\ref{fig:data_mc_ged_af_scale} shows the comparison of the measured and simulated photon transverse momentum,
photon pseudorapidity, photon azimuthal angle,  diphoton invariant mass, diphoton rapidity, and diphoton transverse
momentum distributions. Both the measured yields and kinematic distributions are in accord with the
combination of the LbL scattering signal plus QED $\ee$ and CEP+other background expectations.

\section{Cross section extraction}
\label{sec:sigma_lbyl}

Given the low signal yield available for an extraction of differential cross section
distributions, an integrated fiducial cross section for LbL scattering above a diphoton mass $m^{\gaga} = 5\GeV$ is calculated instead.
The ratio $R$ of cross sections of the light-by-light scattering over the QED $\ee$ processes is measured,
thereby reducing the uncertainties related to trigger and reconstruction efficiencies, and integrated luminosity.
Efficiency uncertainties partially cancel in the ratio, as described later, thanks to a similar selection applied to photons and to electrons;
and the integrated luminosity dependence fully cancels out. The ratio $R$ is defined as
\begin{linenomath*}
\ifthenelse{\boolean{cms@external}}{
\begin{multline}
\label{eq:Req}
 R  =  \frac{\sigma_{\text{fid}} (\gaga \to \gaga)}
 {\sigma(\gaga \to \ee, m^{\ee}>5\GeV)} \\
 = \frac{N^{\gaga,\text{data}}-N^{\gaga,\text{bkg}}}{C^{\gaga}}
 \frac{C^{\Pe\Pe}\,\text{Acc}^{\Pe\Pe}}{N^{\Pe\Pe,\text{data}} \, \mathcal{P}}.
\end{multline}
}
{\begin{equation}
\label{eq:Req}
 R  =  \frac{\sigma_{\text{fid}} (\gaga \to \gaga)}
 {\sigma(\gaga \to \ee, m^{\ee}>5\GeV)}
 = \frac{N^{\gaga,\text{data}}-N^{\gaga,\text{bkg}}}{C^{\gaga}}
 \frac{C^{\Pe\Pe}\,\text{Acc}^{\Pe\Pe}}{N^{\Pe\Pe,\text{data}} \, \mathcal{P}}.
\end{equation}
}
\end{linenomath*}
Here $\sigma_{\text{fid}} (\gaga \to \gaga)$ is the LbL scattering fiducial cross section (\ie\ passing all the aforementioned $\pt$, $\eta$, $m^{\gaga}$
kinematic selection criteria for the single photons and for the photon pair); $\sigma(\gaga \to \ee, m^{\ee}>5\GeV)$ is the total cross section for the QED $\ee$ process
for masses above 5\GeV; $\text{Acc}^{\Pe\Pe} = N^\text{gen} (\pt^\text{gen} > 2\GeV, \abs{\eta^\text{gen}}<2.4, m^{\ee}>5\GeV) / N^\text{gen} (m^{\ee}>5\GeV) = 0.058 \pm 0.001\stat$
is the dielectron acceptance for the fiducial single-electron kinematic selections determined from the \str\ MC generator;
$N^{\gaga,\text{data}}$ is the number of diphoton events passing the selection in data; $N^{\gaga,\text{bkg}}$ is the estimated number of
background events passing all selection criteria; $N^{\Pe\Pe,\text{data}}$ is the number of dielectron events passing
our selection in data; $\mathcal{P}$ is the purity of the estimated fraction of QED $\ee$ signal among these dielectron events; and
$C^{\gaga}$ and $C^{\Pe\Pe}$ are the overall efficiency correction factors, for the $\gaga$ and $\ee$
selections, respectively, that are determined as discussed in the next section.

\subsection{Diphoton analysis efficiencies}

The $C^{\gaga}$ correction factor in Eq.~(\ref{eq:Req}) is obtained through the factorised expression
\begin{linenomath*}
\ifthenelse{\boolean{cms@external}}{
\begin{multline}
  \label{eq:diphoton_corr_fac}
C^{\gaga} = \\
\varepsilon^{\gaga}\, (\text{SF}^{\gamma,\text{reco+ID}})^{2} \, (\text{SF}^{\gaga,\text{trig.}}) \, (\text{SF}^\text{ch.excl.})\, (\text{SF}^\text{neut.excl.}),
\end{multline}
}{
\begin{equation}
C^{\gaga} = \varepsilon^{\gaga}\, (\text{SF}^{\gamma,\text{reco+ID}})^{2} \, (\text{SF}^{\gaga,\text{trig.}}) \, (\text{SF}^\text{ch.excl.})\, (\text{SF}^\text{neut.excl.}),
  \label{eq:diphoton_corr_fac}
\end{equation}
}
\end{linenomath*}
where the diphoton efficiency $\varepsilon^{\gaga}$ is determined using the LbL scattering MC simulation.
This efficiency receives contributions from triggering, photon reconstruction and identification,
and neutral and charged exclusivity criteria that are directly determined from the data via
independent data-to-simulations scale factors, $\text{SF} = \varepsilon^\text{data}/\varepsilon^\text{MC}$,
as explained below.

The diphoton efficiency is first derived from the LbL scattering simulation via:
\begin{linenomath*}
\ifthenelse{\boolean{cms@external}}{
\begin{multline}
\label{eq:diphoton_mc_eff}
 \varepsilon^{\gaga} = \\
 \frac{N^\text{reco} (\et>2\GeV, \abs{\eta^\text{reco}}<2.4, \text{ID, trigger, excl.})}{N^\text{gen} (\et > 2\GeV, \abs{\eta^\text{gen}}<2.4)},
\end{multline}
}{
\begin{equation}
 \varepsilon^{\gaga} = \frac{N^\text{reco} (\et>2\GeV, \abs{\eta^\text{reco}}<2.4, \text{ID, trigger, excl.})}{N^\text{gen} (\et > 2\GeV, \abs{\eta^\text{gen}}<2.4)},
  \label{eq:diphoton_mc_eff}
\end{equation}
}
\end{linenomath*}
where the selection in the numerator and denominator applies to exactly two photons required in each event,
which are also within the fiducial kinematic region in diphoton \pt, mass, and acoplanarity.
It is found to be $\varepsilon^{\gaga} = (20.7 \pm 0.4)\%$, mostly driven by the inefficiencies of the
single photon reconstruction and identification, and of the trigger
($\varepsilon^{\gamma,\text{reco+ID}}, \varepsilon^{\gaga,\text{trig.}}\approx 70\%$).
The quoted uncertainty here is statistical only, reflecting the finite size of the LbL scattering MC sample.

The second term of Eq.~(\ref{eq:diphoton_corr_fac}), the photon reconstruction and identification
efficiency correction $(\text{SF}^{\gamma,\text{reco+ID}})$, is extracted from data by selecting
$\gaga \to \ee(\gamma)$ events, where one of the electrons emits a hard bremsstrahlung
photon due to interaction with the material of the tracker. The \pt of the two electrons in
$\gaga \to \ee$ events being approximately equal, if one of the electrons emits a hard bremsstrahlung
photon, it may not reach the ECAL to be identified as an electron but it can still be reconstructed
in the tracker as a charged particle.

{\tolerance=8000 In a first step, hard-bremsstrahlung events are selected among events passing a trigger requiring
one L1 EG cluster with $\et>5\GeV$, that have exactly two oppositely charged particle tracks
and exactly one electron reconstructed. Among those events, we then look for exactly one photon compatible
with a hard bremsstrahlung, as described below. Such events are used to estimate the efficiency in a tag-and-probe procedure, via
\begin{equation}
\varepsilon^{\gamma,\text{reco+ID, hard-brem}}_\text{data} = \frac{\text{N}^\text{reco+ID,hard-brem}_\text{probe}}{\text{N}^\text{reco+ID, hard-brem}_\text{passing}},
\end{equation}
where denominator and numerator are defined as follows:
\begin{itemize}
\item $\text{N}^\text{reco+ID, hard-brem}_\text{passing}$: Electrons are selected if (i) their direction matches with one
of the two reconstructed tracks within a radius $\Delta R =  \sqrt{\smash[b]{(\Delta \eta)^{2} + (\Delta \phi)^{2}}} < 1.0$
(where $\eta$ and $\phi$ are those of the electron track), (ii) they have \et above 5\GeV, and (iii) their associated ECAL
supercluster is matched within $\Delta R<0.1$ to an L1 EG cluster with $\et>5\GeV$.
The \pt of the track that is not matched with the electron should be below 2\GeV, since we assume that track to
be generated by the electron after bremsstrahlung emission.
The $\pt^{\text{unmatched track}} < 2\GeV$ requirement ensures that this low-\pt charged particle is sufficiently
bent by the magnetic field, and thus the expected photon (extrapolated to the ECAL) and the second electron are sufficiently separated.
Events entering the denominator are not required to have a reconstructed photon.
\item $\text{N}^\text{reco+ID, hard-brem}_\text{probe}$: Events from the denominator are also included in the numerator if
a photon is found with $\et>2\GeV$ that passes the identification criteria.
\end{itemize}

The efficiency is extracted using a fit to the acoplanarity distribution between the electron and the charged-particle track,
and amounts to $\varepsilon^{\gamma,\text{reco+ID, hard-brem}}_\text{data} = (86.5 \pm 7.0)\%$, to be compared with
$\varepsilon_\text{MC}^{\gamma,\text{reco+ID, hard-brem}} = (82.5 \pm 2.0)\%$ in the MC simulation, where uncertainties are statistical (as well as all other uncertainties quoted in this section).
The ratio of these efficiencies is used to define the corresponding
$\text{SF}^{\gamma,\text{reco+ID, hard-brem}} = 1.05 \pm 0.09$ scale factor.
We note that this procedure checks not only the reconstruction and identification efficiency in data,
but also effectively includes bin migrations outside the fiducial $\pt$ range due to the effects of photon energy
scale and resolution. The impact of bin migrations in the final diphoton cross section is found to be below the 1\% level.

Events in the study above comprise exactly two charged-particle tracks, corresponding to the two electrons.
They do not probe the possibility that the photon, reconstructed in the ECAL, has previously also interacted in the tracker
generating an $\Pep\Pem$ pair that has been also reconstructed as one or two additional displaced low-\pt charged-particle tracks.
In the case of an LbL scattering event, such a genuine signal event would be discarded by the strict charged exclusivity criterion,
which is applied independently of the proximity of the tracks to the photon, in order to keep the QED $\ee$ background to a minimum.
The modeling of this efficiency loss in simulation is checked using hard-bremsstrahlung events with a similar selection as above,
except that up to two additional charged-particle
tracks are now allowed in the event. We check the fraction of events where no additional track, more displaced than the one with $\pt<2\GeV$ required in the selection,
is found in a window $\abs{\Delta\eta}<0.15$, $\abs{\Delta\phi}<0.7$ around the photon. This efficiency amounts to $\varepsilon^{\gamma,\text{tk veto}}_\text{data} = (89.9 \pm 1.7)\%$ in data,
and $\varepsilon^{\gamma,\text{tk veto}}_\text{MC} = (91.1 \pm 1.2)\%$ in the MC QED $\ee$ simulation. The ratio of these efficiencies gives
$\text{SF}^{\gamma,\text{tk veto}} = 0.99 \pm 0.02$.
The final overall scale factor for reconstruction and identification, accounting for the modeling of photon conversions in the
MC simulation and the efficiency to reconstruct the associated displaced tracks, is then
$\text{SF}^{\gamma,\text{reco+ID}} = \text{SF}^{\gamma,\text{reco+ID, hard-brem}} \, \text{SF}^{\gamma,\text{tk veto}} = 1.04 \pm 0.09$.

The third term of Eq.~(\ref{eq:diphoton_corr_fac}) accounts for the trigger selection efficiency. Exclusive diphoton
events are selected using an L1 trigger requiring two electromagnetic clusters with $\et>2\GeV$, and no
activity (above noise thresholds) in at least one of the HF calorimeters.
These two components of the trigger, the electromagnetic cluster selection and the HF energy veto, are verified independently in data.
The efficiency for reconstructing an L1 EG cluster with $\et>2\GeV$ is verified using a tag-and-probe technique on QED $\ee$ events, where the dielectron
acoplanarity is fitted to extract the signal and measure the efficiency. The same selection criteria used in the main analysis
are applied, including the exclusivity requirements. Events are further selected using a supporting trigger requiring one L1 EG cluster
with $\et>5\GeV$ with the same HF energy veto as the analysis trigger.
The L1 EG cluster used in the trigger is matched (using the same criterion mentioned above) to one of the two electrons reconstructed offline, called the tag.
The other electron in the event is the probe, and it qualifies as a passing probe if it is matched to an
L1 EG cluster with $\et>2\GeV$. The efficiency is then the fraction of probes that are also passing probes, and it is in the 45--100\%
range, with the lowest efficiency found close to the $\et = 2\GeV$ threshold and at high $|\eta|$.
Scale factors are determined from the data-MC differences, as a function of \et, in two
$\abs{\eta}$ bins. Applying them to the LbL simulation, we find an integrated scale factor of
$1.12 \pm 0.31\stat$. The same QED $\ee$ sample is used to test the HF veto component of the analysis trigger.
This time, we apply the nominal dielectron selection, including exclusivity requirements, but for a data sample
collected with a trigger requiring a single-EG object in the HLT with $\et>10\GeV$ and $\abs{\eta}<1.5$ plus
a small amount of energy in the HFs corresponding to about 50\% of the most peripheral PbPb events.
Both electrons in the event are then matched to an L1 EG cluster with $\et>2\GeV$.
We find that $(100^{+0}_{-3})\%$ of the selected events also pass the analysis trigger, \ie\
satisfy the HF veto in the trigger, in perfect agreement with the result
predicted from the MC simulation. Combining the results of the studies above, the scale factor
$\text{SF}^{\gaga,\text{trig.}} = 1.12 \pm 0.31$ is obtained for the ratio of the product of
trigger efficiencies in data to that obtained from the MC simulation.

The last two terms of Eq.~(\ref{eq:diphoton_corr_fac}) account for the efficiency of the exclusivity selections.
The fraction of events passing the QED dielectron selection, with the exception of the charged and neutral exclusivity criteria,
are analysed. Using the acoplanarity distribution to extract the signal, we find that  $(92.5 \pm 0.3)\%$ of the events feature
no additional track in the event, to be compared to $(99.3 \pm 0.1)\%$ in simulation. We deduce that the corresponding
scale factor is $\text{SF}^{\text{ch.excl.}} = 0.93 \pm 0.01$.
A similar strategy is used for the neutral exclusivity selection, this time in events passing the corresponding requirements.
This efficiency is found to be $(89.9 \pm 1.4)\%$ in data, and $(96.9 \pm 1.3)\%$ in simulation. This scale factor is then
$\text{SF}^{\text{neut.excl.}} = 0.93 \pm 0.02$. Differences between the exclusivity efficiencies in data and MC simulation
are likely due to the presence of nonexclusive events, such as $\gamma\gamma\to\ee$ processes with a small hadronic
overlap of the lead ions, whose modeling is currently not available in the Monte Carlo generators.
The incorporation of such nonexclusive events in the definition of the signal is irrelevant, 
because both $\text{SF}^{\text{ch.excl.}}$ and $\text{SF}^{\text{neut.excl.}}$ cancel out in the R ratio, as explained in Section~\ref{sec:syst}.

\subsection{Dielectron analysis efficiencies}

For the exclusive dielectron analysis, the efficiency is estimated using the \str\ MC simulation via
\begin{linenomath*}
\ifthenelse{\boolean{cms@external}}{
\begin{multline}
  \label{eq:diele_mc_eff}
 \varepsilon^{\Pe\Pe}  = \\
  \frac{N^\text{reco} (\pt^\text{reco} > 2\GeV, \abs{\eta^\text{reco}}<2.4, \text{ID, trigger, excl.})}{N^\text{gen} (\pt^\text{gen} > 2\GeV, \abs{\eta^\text{gen}}<2.4)},
\end{multline}
}{
\begin{equation}
 \varepsilon^{\Pe\Pe}  =  \frac{N^\text{reco} (\pt^\text{reco} > 2\GeV, \abs{\eta^\text{reco}}<2.4, \text{ID, trigger, excl.})}{N^\text{gen} (\pt^\text{gen} > 2\GeV, \abs{\eta^\text{gen}}<2.4)},
  \label{eq:diele_mc_eff}
\end{equation}
}
\end{linenomath*}
where the kinematic criteria in the numerator and denominator are applied to exactly the two electrons required in the event.
The different components of the electron efficiency are again checked using data, via a factorised expression
for the corresponding correction factors:
\begin{linenomath*}
\ifthenelse{\boolean{cms@external}}{
\begin{multline}
  \label{eq:diele_global_eff}
 C^{\Pe\Pe} = \\
\varepsilon^{\Pe\Pe}\, (\text{SF}^{\Pe,\text{reco+ID}})^{2} \, (\text{SF}^{\Pe\Pe,\text{trig.}}) \, (\text{SF}^\text{ch.excl.})\, (\text{SF}^\text{neutral excl.}).
\end{multline}
}{
\begin{equation}
 C^{\Pe\Pe} = \varepsilon^{\Pe\Pe}\, (\text{SF}^{\Pe,\text{reco+ID}})^{2} \, (\text{SF}^{\Pe\Pe,\text{trig.}}) \, (\text{SF}^\text{ch.excl.})\, (\text{SF}^\text{neutral excl.}).
  \label{eq:diele_global_eff}
\end{equation}
}
\end{linenomath*}
Most of the scale factors are common with those used in the diphoton analysis
(since they are computed using the larger statistical sample of electrons in data),
except for the reconstruction and identification efficiency, which we check again for electrons.
For the latter, a tag-and-probe technique using a fit to the acoplanarity distribution in QED $\ee$ events is used,
as done for the diphoton case, except that now the probe is a charged-particle track that is a
passing probe if it is matched to an electron
passing the reconstruction and identification criteria. We find an efficiency of $(89.4 \pm 1.2)\%$ in data, consistent with
$(90.4 \pm 1.3)\%$ in the MC simulation, corresponding to a scale factor of $\text{SF}^{\text{\Pe, reco+ID}} = 0.99 \pm 0.02$.

The scale factor for the trigger efficiency is also recomputed using the \pt spectrum in the QED $\ee$ MC simulation,
using the same \pt- and $\abs{\eta}$-dependent scale factors as for $\text{SF}^{\gaga,\text{trig.}}$,
leading to $\text{SF}^{\Pe\Pe,\text{trig.}} = 1.09 \pm 0.16$.

\subsection{Summary of the efficiencies}

The overall cross section measurement efficiencies, efficiencies in simulation, as well as
the individual data-to-simulation scale factors, obtained for the diphoton and dielectron
analyses are summarised in Table~\ref{tab:eff_summary}. 
Since the data-to-simulation scale factors are consistent with unity, they are not included 
in the numbers listed in Table~\ref{tab:event_sel_ged} nor in the results plotted in 
Figs.~\ref{fig:qed_ee_acop}--\ref{fig:data_mc_ged_af_scale}, but they are used to obtain the results in Section~\ref{sec:results}.
The overall diphoton cross section
efficiency, Eq.~(\ref{eq:diphoton_mc_eff}), is about 20\% compared with about 10\% for dielectrons, Eq.~(\ref{eq:diele_mc_eff}).
The dielectron analysis is a factor of two less efficient than the diphoton one, because each single electron
has a relatively larger probability of losing energy by bremsstrahlung before reaching the ECAL, and
therefore their probability to pass the trigger selection threshold and/or their energy be properly reconstructed is smaller.
Such efficiency losses are further enhanced as they enter squared for two electrons to pass the trigger or be
concurrently reconstructed above the \pt and mass thresholds.

\begin{table*}[htbp]
 \topcaption{\label{tab:eff_summary}
  Summary of the overall cross section measurement efficiencies $C^{\gaga,\Pe\Pe}$, efficiencies from simulation $\varepsilon^{\gaga,\Pe\Pe}$,
  and individual data-to-simulation scale factors $\text{SF}^{\gaga,\Pe\Pe}$, obtained for the diphoton and dielectron analyses. ``Reco. and ID'' stands for
  reconstruction and identification. All quoted uncertainties are systematic.}
 \centering
 \begin{tabular}{lrcl}
  \hline
  Diphoton global efficiency, Eq.~(\ref{eq:diphoton_corr_fac}) & $C^{\gaga}$ & $=$& $(21.5 \pm 6.5)\%$  \\
  Diphoton efficiency (from simulation) & $\varepsilon^{\gaga}$ & $=$ & $(20.7 \pm 0.4)\%$ \\
  $\gamma$ reco. and ID data-to-simulation scale factor& $\text{SF}^{\gamma, \text{reco+ID}}$ & $=$ & $ 1.04 \pm 0.09$ \\
  Diphoton trigger selection  data-to-simulation scale factor& $\text{SF}^{\gaga, \text{trig.}}$ & $=$ & $ 1.12 \pm 0.31$ \\[\cmsTabSkip]
  Dielectron global efficiency, Eq.~(\ref{eq:diele_global_eff}) & $C^{\Pe\Pe}$ & $=$ & $(9.4 \pm 1.5)\%$  \\
  Dielectron efficiency (simulation) & $\varepsilon^{\Pe\Pe}$ & $=$ & $ (10.4 \pm 0.1)\%$ \\
  $\Pe^\pm$ reco. and ID data-to-simulation scale factor& $\text{SF}^{\text{\Pe, reco+ID}}$ & $=$ & $ 0.98 \pm 0.04$ \\
  Dielectron trigger selection data-to-simulation scale factor & $\text{SF}^{\text{\Pe\Pe,trig.}}$ & $=$ & $ 1.09 \pm 0.16$ \\[\cmsTabSkip]
  Charged exclusivity data-to-simulation scale factor & $\text{SF}^{\text{ch.excl.}}$ & $=$ & $ 0.93 \pm 0.01$ \\
  Neutral exclusivity data-to-simulation scale factor & $\text{SF}^{\text{neut.excl.}}$ & $=$ & $ 0.93 \pm 0.02$ \\\hline
 \end{tabular}
\end{table*}

\section{Systematic uncertainties}
\label{sec:syst}

The main sources of uncertainty in the LbL scattering and QED $\ee$ production measurements are related to the trigger
and single $\gamma,\Pe^\pm$ reconstruction efficiencies (Table~\ref{tab:eff_summary}).
The uncertainty in the latter is doubled in the total cross section, since we consider diphoton and dielectron final states.
No additional uncertainty in the photon energy scale and resolution is considered, since possible data-simulation
differences are already included in the derivation of the reconstruction and identification scale factors.
Systematic uncertainties have been estimated for the different terms defining the ratio $R$ of the LbL scattering
over QED $\ee$ production cross sections given by Eq.~(\ref{eq:Req}), and are summarised in Table~\ref{tab:syst_summary}.
Because the scale factors used for the trigger efficiency are common to the diphoton and dielectron analyses,
their associated uncertainty cancels partially in the ratio. However, because of different reconstruction and identification
efficiencies, the \et spectrum of photons is different from that of the electrons, leading to only incomplete cancellation of the
uncertainty. Assuming the uncertainty in each individual \pt-binned scale factor is fully correlated between the photon and electron cases, 
but with no correlation between the scale factors for different \pt, we propagate these uncertainties simultaneously to the numerator and 
denominator of the ratio $\text{SF}^{\gaga,\text{trigger}}/\text{SF}^{\Pe\Pe,\text{trigger}}$, resulting in a 12\% uncertainty in that ratio.
The charged and neutral exclusivity scale factors, common to the diphoton and dielectron measurements,
are assumed to cancel in the ratio $R$.
The rest of the SF terms listed in Table~\ref{tab:eff_summary} have small statistical uncertainties from the finite
size of the MC samples used to derive them, which propagate into percent uncertainties in the final cross section,
and are neglected here.

Among the other parameters in Eq.~(\ref{eq:Req}), the normalisation of the CEP and QED $\ee$ backgrounds in the signal
region propagates into a $16\%$ uncertainty in the background yield (resulting in an $6\%$ uncertainty in the cross section measurement),
accounting for the finite size of the MC samples. An additional uncertainty of $25\%$ ($10\%$ in the final cross section),
reflecting the finite size of the data sample at high acoplanarities used for the absolute normalisation of
the CEP plus residual nonexclusive backgrounds, is considered as a statistical uncertainty rather than a systematic one.

The final systematic uncertainty is obtained from adding in quadrature the individual uncertainties
and is listed in the last row of Table~\ref{tab:syst_summary}.

\begin{table*}[htbp]
 \topcaption{\label{tab:syst_summary}Summary of the systematic uncertainties in the ratio of the fiducial LbL  scattering
  to total QED $\ee$ cross sections.}
 \centering
  \begin{tabular}{lc}
  \hline
   Photon reconstruction and identification ($\text{SF}^{\gamma, \text{reco+ID}}$) &  $(2{\times}9)\%$ \\
   Electron reconstruction and identification ($\text{SF}^{\text{\Pe, reco+ID}}$)  &  $(2{\times}2.5)\%$ \\
   Trigger & 12\% \\
   Size of simulated background samples & 6\% \\[\cmsTabSkip]
   Total & 23\% \\
   \hline
  \end{tabular}
\end{table*}

\section{Results} \label{sec:results}
\subsection{Light-by-light cross section}
\label{sec:xsec}

The compatibility of the data with the background-only hypothesis has been evaluated from the measured acoplanarity
distribution (Fig.~\ref{fig:aco_plot}), using a profile-likelihood ratio as a test statistic, including all systematic uncertainties as nuisance
parameters with log-normal priors~\cite{Barlow:1993dm,Conway:2011in}. The uncertainty due to the finite size of the MC samples
is also included as an additional nuisance parameter for each bin of the histogram. The significance of the excess at low diphoton
acoplanarity in data, estimated from the expected distribution of the test statistic for the background-only hypothesis obtained
with pseudo-experiments, is 3.7 standard deviations (3.5 standard deviations expected). If using only the total number of events
observed and expected in the region $\Aco < 0.01$, we obtain a significance of 3.4 standard deviations (3.2 expected).

The final ratio of the fiducial LbL scattering to the total QED $\ee$ cross sections is obtained from Eq.~(\ref{eq:Req}), and amounts to
\begin{linenomath*}
\begin{equation}
R = (25.0 \pm9.6\stat \pm 5.8\syst) \times 10^{-6},
\label{eq:R_final}
\end{equation}
\end{linenomath*}
where the statistical uncertainty includes the normalisation uncertainties of the CEP and QED backgrounds, added in quadrature.
The fiducial cross section is obtained from the theoretical prediction of $\sigma(\gaga \to \ee, m^{\Pe\Pe}>5\GeV) = 4.82 \pm
0.48\thy\unit{mb}$ from \str, where the 10\% uncertainty is derived from alternative approaches~\cite{Harland-Lang:2018iur} to compute the nonhadronic-overlap condition in the simulation:
\begin{linenomath*}
\begin{equation}
\sigma_\text{fid} (\gaga \to \gaga) = 120 \pm46\stat \pm28\syst\pm12\thy\unit{nb},
\end{equation}
\end{linenomath*}
in good agreement with the theoretical LbL prediction~\cite{d'Enterria:2013yra} in the fiducial region,
defined in Section~\ref{sec:sigma_lbyl}, of
\begin{linenomath*}
\begin{equation}
\sigma_\text{fid} (\gaga \to \gaga) = 116 \pm 12\unit{nb}.
\end{equation}
\end{linenomath*}
The 10\% uncertainty in the LbL theoretical prediction covers different implementations of the nonhadronic-overlap
condition computed with a Glauber model~\cite{Loizides:2017ack} for varying Pb radius and
nucleon-nucleon cross section values, as well as neglected NLO corrections.

\subsection{Exclusion limits on axion-like particle production}
The measured invariant mass distribution (Fig.~\ref{fig:data_mc_ged_af_scale}, center right) is used to search for possible narrow diphoton
resonances, such as pseudoscalar axion-like particles produced in the process $\gaga\to\Pa\to\gaga$~\cite{Knapen:2016moh}.
The LbL, QED, and CEP+other continuum processes are considered as backgrounds in this search. Fully simulated \textsc{starlight} samples for
various ALP masses, $m_\Pa$, ranging from 5 to 90\GeV are reconstructed with the same code used for the LbL analysis in order to estimate
the ALP acceptance and efficiency, as well as the expected reconstructed diphoton mass template distributions.
Corrections to the efficiency estimated in the MC simulation are derived based on data, and applied in the same way as for the LbL analysis.
A binned maximum likelihood fit of the signal and background contributions is performed on the data,
where systematic uncertainties are included as nuisance parameters with a log-normal prior.
The \CLs criterion~\cite{CLS2,CLS1}, with a profile likelihood ratio as test statistic~\cite{ATLAS:2011tau},
is used to extract exclusion limits
in the $\sigma(\gaga \to \Pa \to \gaga)$ cross section at 95\% confidence level (\CL).
Limits on $\sigma(\gaga \to \Pa \to \gaga)$ cross section for axion-like particles with masses
5--90\GeV are set in the  1500--20\unit{nb} range (Fig.~\ref{fig:axion_limits}). The 68 and 95\% \CL bands around the
expected limits are obtained using pseudo-experiments.

\begin{figure*}[htbp]
\centering
\includegraphics[width=0.6\textwidth]{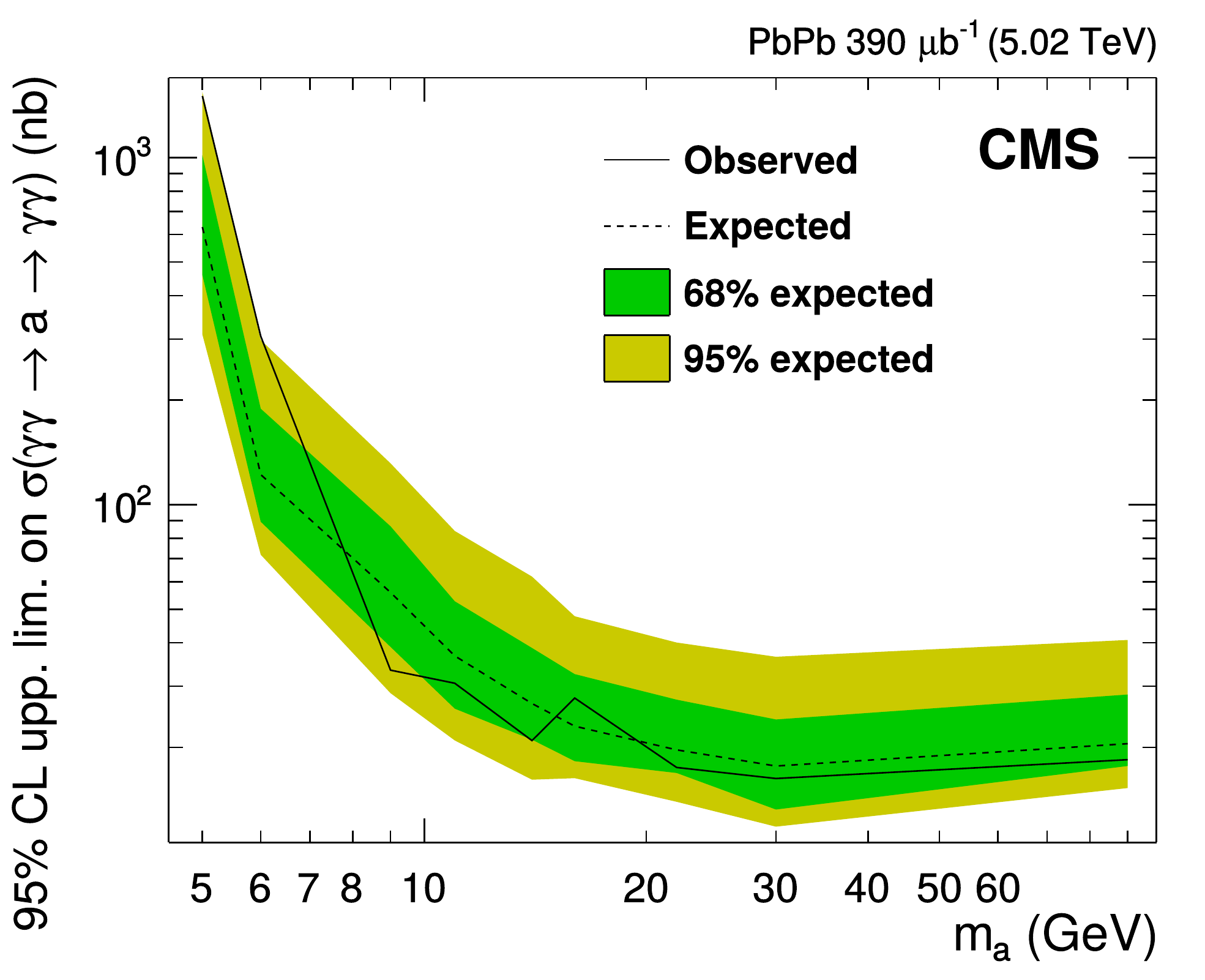}
\caption{\label{fig:axion_limits}
Observed (full line) and expected (dotted line) 95\% \CL limits on the production cross section $\sigma(\gaga \to \Pa \to \gaga)$
as a function of the ALP mass $m_\Pa$ in ultraperipheral PbPb collisions at $\sqrtsNN = 5.02\TeV$.
The inner (green )and outer (yellow) bands indicate the regions containing 68 and 95\%, respectively, of the distribution
of limits expected under the background-only hypothesis.}
\end{figure*}

The cross section limits shown in Fig.~\ref{fig:axion_limits} are used to set exclusion limits in
the $g_{\Pa\gamma}$ vs, $m_\Pa$ plane, where $g_{\Pa\gamma}\equiv 1/\Lambda$ is the ALP coupling to photons (with $\Lambda$ being
the energy scale associated with the underlying U(1) symmetry whose spontaneous breaking generates the ALP mass). Two scenarios are considered where
the ALP couples to photons $F^{\mu\nu}$ alone, or also to hypercharge $B^{\mu\nu}$ with operators: $\Pa F\widetilde{F}/4\Lambda$ and
$\Pa B\widetilde{B}/(4\Lambda \cos^2\theta_\textrm{W})$ (where $\theta_\textrm{W}$ is the Weinberg angle), respectively~\cite{Knapen:2016moh}.
The derived constraints on the ALP mass and its coupling to photons are compared in Fig.~\ref{fig:axion_aFFBB} to those
obtained~\cite{Knapen:2016moh,limits_lep} from various experiments~\cite{Chatrchyan:2012tv,limits_opal,limits_atlas_2gamma,limits_atlas_3gamma},
assuming  a 100\% ALP decay branching fraction to diphotons.
For an ALP sensitive to the electromagnetic current alone (left plot), our exclusion limits are the best so far over the $m_\Pa = 5$--50\GeV mass range.
In the case of extra ALP couplings to electroweak currents (right plot), our result provides new constraints in the
$m_\Pa = 5$--10\GeV region.

\begin{figure*}[hbtp]
\centering
\includegraphics[width=0.49\textwidth]{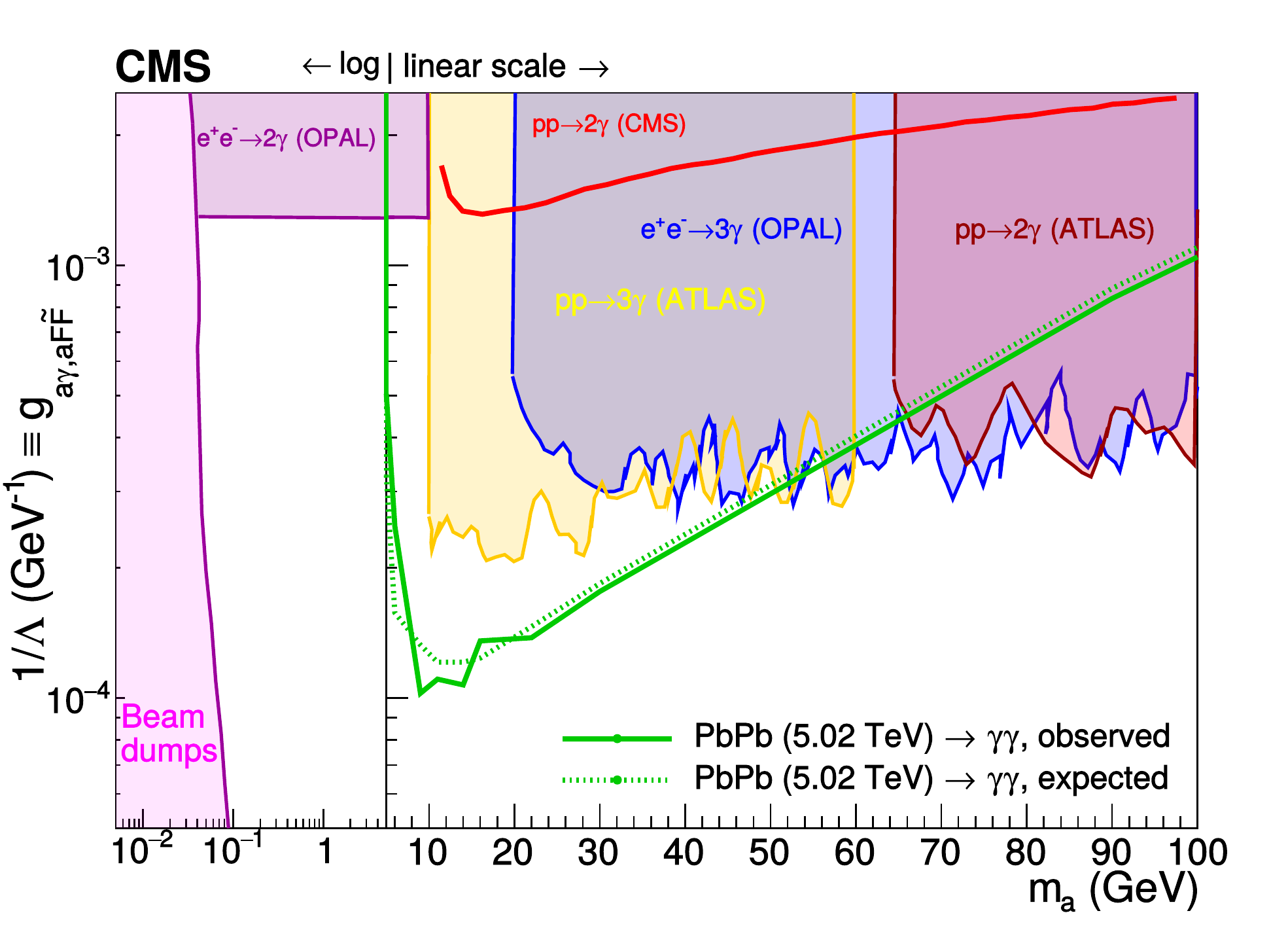}\hfill
\includegraphics[width=0.49\textwidth]{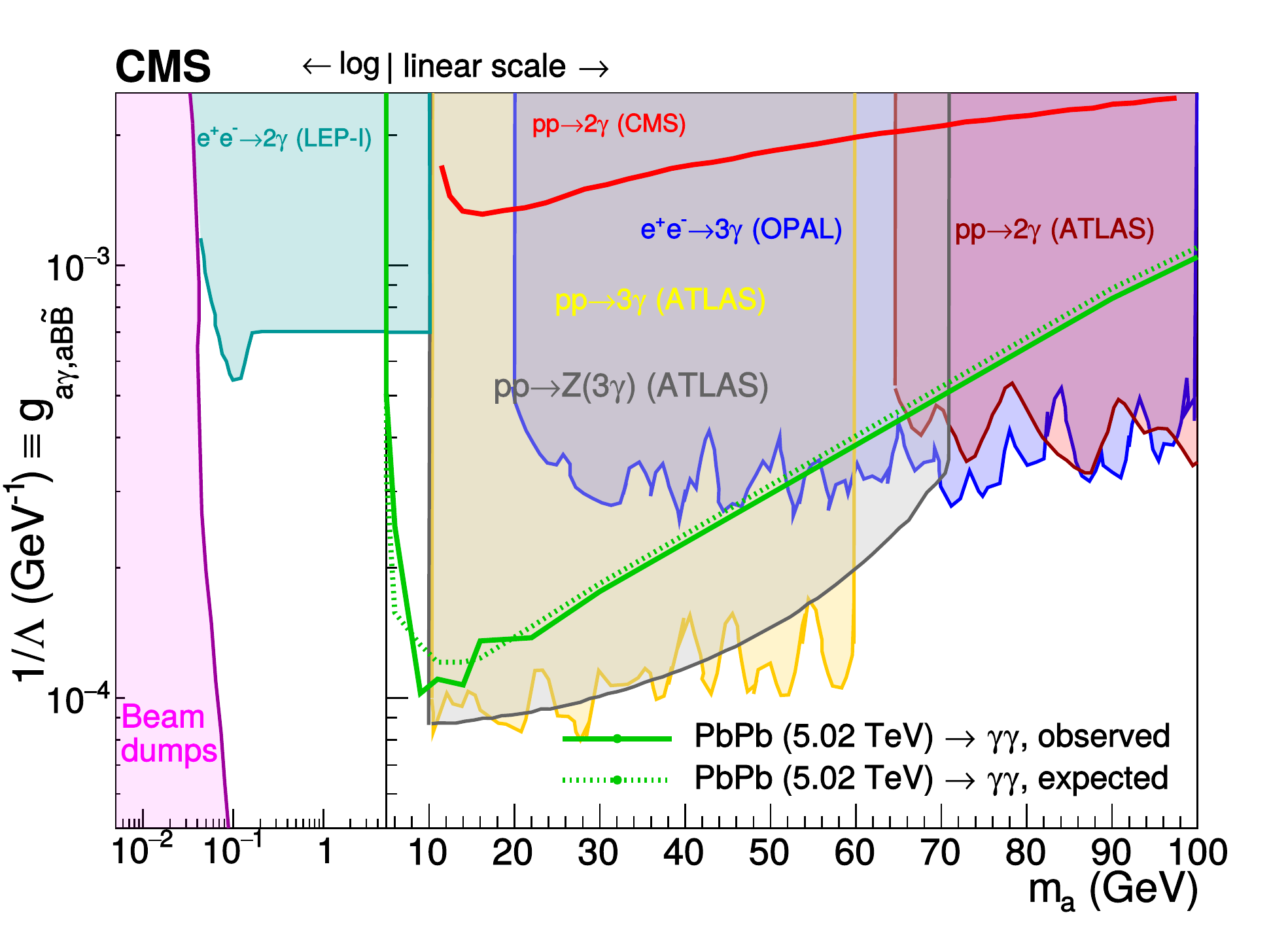}
\caption{Exclusion limits at 95\% \CL in the ALP-photon coupling $g_{\Pa\gamma}$ versus ALP
mass $m_\Pa$ plane, for the operators $\Pa F\widetilde{F}/4\Lambda$ (left, assuming
ALP coupling to photons only) and $\Pa B\widetilde{B}/(4\Lambda \cos^2\theta_\textrm{W})$ (right,
including also the hypercharge coupling, thus processes involving the {\PZ} boson) derived in Refs.~\cite{Knapen:2016moh,limits_lep}
from measurements at beam dumps~\cite{Dobrich:2015jyk}, in \Pep\Pem\ collisions at LEP-I~\cite{limits_lep} and LEP-II~\cite{limits_opal},
and in \Pp\Pp collisions at the LHC~\cite{Chatrchyan:2012tv,limits_atlas_2gamma,limits_atlas_3gamma}, and compared to the present PbPb limits.}
\label{fig:axion_aFFBB}
\end{figure*}

\section{Summary}

Evidence for light-by-light (LbL) scattering, $\gaga \to \gaga$, in ultraperipheral PbPb collisions
at a centre-of-mass energy per nucleon pair of 5.02\TeV has been reported, based on a data sample corresponding to
an integrated luminosity of 390\mubinv recorded by the CMS experiment at the LHC in 2015. Fourteen LbL-scattering candidate
events passing all selection requirements have been observed, with photon transverse energy above $2\GeV$ and pseudorapidity
$\abs{\eta} < 2.4$, diphoton invariant mass greater than 5\GeV, diphoton transverse momentum lower than 1\GeV, and
diphoton acoplanarity below 0.01. Both the measured total yields and kinematic distributions are in accord
with the expectations for the LbL scattering signal plus small residual backgrounds that are mostly
from misidentified exclusive dielectron ($\gaga\to\ee$) and gluon-induced central exclusive ($\Pg\Pg\to \gaga$) processes.
The observed (expected) significance of the LbL scattering signal over the background-only
expectation is  3.7 (3.5) standard deviations.
The ratio of the fiducial LbL scattering to the total QED dielectron cross sections is $R = (25.0 \pm9.6\stat \pm 5.8\syst) \times 10^{-6}$.
From the theoretical $\gaga \to \ee$ cross section prediction, we derive a fiducial light-by-light scattering cross section,
$\sigma_\text{fid} (\gaga \to \gaga) = 120 \pm46\stat \pm28\syst\pm12\thy\unit{nb}$,
consistent with the standard model expectation. The measured exclusive diphoton invariant mass distribution is
used to set new exclusion limits on the production of pseudoscalar axion-like particles (ALPs), via the process $\gaga \to \Pa \to \gaga$, over
the $m_\Pa = 5$--90\GeV mass range. For ALPs coupling to the electromagnetic (and electroweak) current,
the derived exclusion limits are currently the best over the $m_\Pa = 5$--50\GeV (5--10\GeV) mass range.

\begin{acknowledgments}

We congratulate our colleagues in the CERN accelerator departments for the excellent performance of the LHC and thank the technical and administrative staffs at CERN and at other CMS institutes for their contributions to the success of the CMS effort. In addition, we gratefully acknowledge the computing centres and personnel of the Worldwide LHC Computing Grid for delivering so effectively the computing infrastructure essential to our analyses. Finally, we acknowledge the enduring support for the construction and operation of the LHC and the CMS detector provided by the following funding agencies: BMBWF and FWF (Austria); FNRS and FWO (Belgium); CNPq, CAPES, FAPERJ, FAPERGS, and FAPESP (Brazil); MES (Bulgaria); CERN; CAS, MoST, and NSFC (China); COLCIENCIAS (Colombia); MSES and CSF (Croatia); RPF (Cyprus); SENESCYT (Ecuador); MoER, ERC IUT, and ERDF (Estonia); Academy of Finland, MEC, and HIP (Finland); CEA and CNRS/IN2P3 (France); BMBF, DFG, and HGF (Germany); GSRT (Greece); NKFIA (Hungary); DAE and DST (India); IPM (Iran); SFI (Ireland); INFN (Italy); MSIP and NRF (Republic of Korea); MES (Latvia); LAS (Lithuania); MOE and UM (Malaysia); BUAP, CINVESTAV, CONACYT, LNS, SEP, and UASLP-FAI (Mexico); MOS (Montenegro); MBIE (New Zealand); PAEC (Pakistan); MSHE and NSC (Poland); FCT (Portugal); JINR (Dubna); MON, RosAtom, RAS, RFBR, and NRC KI (Russia); MESTD (Serbia); SEIDI, CPAN, PCTI, and FEDER (Spain); MOSTR (Sri Lanka); Swiss Funding Agencies (Switzerland); MST (Taipei); ThEPCenter, IPST, STAR, and NSTDA (Thailand); TUBITAK and TAEK (Turkey); NASU and SFFR (Ukraine); STFC (United Kingdom); DOE and NSF (USA).

\hyphenation{Rachada-pisek} Individuals have received support from the Marie-Curie programme and the European Research Council and Horizon 2020 Grant, contract No. 675440 (European Union); the Leventis Foundation; the A. P. Sloan Foundation; the Alexander von Humboldt Foundation; the Belgian Federal Science Policy Office; the Fonds pour la Formation \`a la Recherche dans l'Industrie et dans l'Agriculture (FRIA-Belgium); the Agentschap voor Innovatie door Wetenschap en Technologie (IWT-Belgium); the F.R.S.-FNRS and FWO (Belgium) under the ``Excellence of Science - EOS" - be.h project n. 30820817; the Ministry of Education, Youth and Sports (MEYS) of the Czech Republic; the Lend\"ulet (``Momentum") Programme and the J\'anos Bolyai Research Scholarship of the Hungarian Academy of Sciences, the New National Excellence Program \'UNKP, the NKFIA research grants 123842, 123959, 124845, 124850 and 125105 (Hungary); the Council of Science and Industrial Research, India; the HOMING PLUS programme of the Foundation for Polish Science, cofinanced from European Union, Regional Development Fund, the Mobility Plus programme of the Ministry of Science and Higher Education, the National Science Center (Poland), contracts Harmonia 2014/14/M/ST2/00428, Opus 2014/13/B/ST2/02543, 2014/15/B/ST2/03998, and 2015/19/B/ST2/02861, Sonata-bis 2012/07/E/ST2/01406; the National Priorities Research Program by Qatar National Research Fund; the Programa Estatal de Fomento de la Investigaci{\'o}n Cient{\'i}fica y T{\'e}cnica de Excelencia Mar\'{\i}a de Maeztu, grant MDM-2015-0509 and the Programa Severo Ochoa del Principado de Asturias; the Thalis and Aristeia programmes cofinanced by EU-ESF and the Greek NSRF; the Rachadapisek Sompot Fund for Postdoctoral Fellowship, Chulalongkorn University and the Chulalongkorn Academic into Its 2nd Century Project Advancement Project (Thailand); the Welch Foundation, contract C-1845; and the Weston Havens Foundation (USA).

\end{acknowledgments}

\bibliography{auto_generated}
\cleardoublepage \appendix\section{The CMS Collaboration \label{app:collab}}\begin{sloppypar}\hyphenpenalty=5000\widowpenalty=500\clubpenalty=5000\vskip\cmsinstskip
\textbf{Yerevan Physics Institute, Yerevan, Armenia}\\*[0pt]
A.M.~Sirunyan, A.~Tumasyan
\vskip\cmsinstskip
\textbf{Institut f\"{u}r Hochenergiephysik, Wien, Austria}\\*[0pt]
W.~Adam, F.~Ambrogi, E.~Asilar, T.~Bergauer, J.~Brandstetter, M.~Dragicevic, J.~Er\"{o}, A.~Escalante~Del~Valle, M.~Flechl, R.~Fr\"{u}hwirth\cmsAuthorMark{1}, V.M.~Ghete, J.~Hrubec, M.~Jeitler\cmsAuthorMark{1}, N.~Krammer, I.~Kr\"{a}tschmer, D.~Liko, T.~Madlener, I.~Mikulec, N.~Rad, H.~Rohringer, J.~Schieck\cmsAuthorMark{1}, R.~Sch\"{o}fbeck, M.~Spanring, D.~Spitzbart, A.~Taurok, W.~Waltenberger, J.~Wittmann, C.-E.~Wulz\cmsAuthorMark{1}, M.~Zarucki
% ...
\vskip\cmsinstskip
% ...
1:  Also at Vienna University of Technology, Vienna, Austria\\
% ...
\end{sloppypar}
\end{document}